\documentclass[useAMS,usedcolumn]{mn2e}
\usepackage{amssymb}
\usepackage{times}
\usepackage{lscape}
\usepackage{graphicx}
\usepackage{pifont}
\usepackage{amsfonts}
\usepackage{amsmath}
\usepackage{rotating}
\usepackage{stfloats}
\usepackage{longtable}
\usepackage{rotating}
\usepackage{cleveref}							
\usepackage{url}									
\usepackage{latexsym}
\usepackage{enumerate}

\def\ha{H$\alpha$}

\def\NII{[N\,\textsc{ii}]}

\def\OIII{[O\,\textsc{iii}]}

\def\HII{H\,\textsc{ii}}

\def\p0{\phantom{0}}

\def\cm3{cm$^{-3}$}

\def\12{$^{12}$CO}
\def\13co{$^{13}$CO}

\def\ha{H$\alpha$}

\def\HII{H\,{\sc ii}}

\def\la{{\hbox{$\lesssim$}}}

\def\arcdeg{\hbox{$^\circ$}}
\def\arcmin{\hbox{$^\prime$}}
\def\arcsec{\hbox{$^{\prime\prime}$}}
\def\ergcms{erg\,cm$^{-2}$\,s$^{-1}$}

\def\ha{H$\alpha$}

\def\lessim{\raise-.5ex\hbox{$\buildrel<\over{\scriptstyle\mathtt{\sim}}$}}
\def\grtsim{\raise-.5ex\hbox{$\buildrel>\over{\scriptstyle\mathtt{\sim}}$}}

\begin{document}
\title[Flux calibration of the SuperCOSMOS H$\alpha$ Survey]{Flux calibration of the AAO/UKST SuperCOSMOS H$\alpha$ Survey}

\author[D.J. Frew et al.]
{David J. Frew$^{1,2}$\thanks{E-mail: david.frew@mq.edu.au}, 
Ivan S. Boji\v{c}i\'c$^{1,2,3}$, Quentin A. Parker$^{1,2,3}$, Mark J. Pierce$^{1,4}$, \newauthor  M.L.P. Gunawardhana$^{5}$ and W.A. Reid$^{1,2}$\\
$^{1}$ Department of Physics and Astronomy, Macquarie University, Sydney, NSW 2109 Australia\\
$^{2}$ Research Centre for Astronomy, Astrophysics and Astrophotonics, Macquarie University, Sydney, NSW 2109 Australia\\
$^{3}$ Australian Astronomical Observatory, PO Box 915, North Ryde, NSW 1670, Australia\\
$^{4}$ Department of Physics, University of Bristol, Bristol, BS8 1TL, UK\\
$^{5}$ Sydney Institute for Astronomy, School of Physics, The University of Sydney, NSW 2006, Australia\\
}
 
\date{ Accepted . Received ; in original form}       

\maketitle

\begin{abstract}
The AAO/UKST SuperCOSMOS H$\alpha$ Survey (SHS) was, when completed in 2003, a powerful addition to extant wide-field surveys.  The combination of areal coverage, spatial resolution and sensitivity in a narrow imaging band, still marks it out today as an excellent resource for the astronomical community.  The 233 separate fields are available online in digital form, with each field covering 25 square degrees.  The SHS has been the motivation for equivalent surveys in the north, and new digital H$\alpha$ surveys now beginning in the south such as VPHAS+.  It has been the foundation of  many important discovery projects with the Macquarie/AAO/Strasbourg H$\alpha$ planetary nebula project being a particularly successful example.  However, the full potential of the SHS has been hampered by lack of a clear route to acceptable flux calibration from the base photographic data. We have determined the calibration factors for 170 individual SHS fields, and present a direct pathway to the measurement of integrated H$\alpha$ fluxes and surface brightnesses for resolved nebulae detected in the SHS.  We also include a catalogue of integrated \ha\ fluxes for $>$100 planetary and other nebulae measured from the SHS, and use these data to show that fluxes, accurate to $\pm$\,0.10\,--\,0.14\,dex ($\sim$25--35 per cent), can be obtained from these fields.  For the remaining 63 fields, a mean calibration factor of 12.0\,counts\,pix$^{-1}$\,R$^{-1}$ can be used, allowing the determination of reasonable integrated fluxes accurate to better than $\pm$0.2\,dex ($\sim$50 per cent).  We outline the procedures involved and the caveats that need to be appreciated in achieving such flux measurements.  This paper forms a handy reference source that will significantly increase the scientific utility of the SHS.  
\end{abstract}
\section{Introduction}
The AAO/UKST SuperCOSMOS H$\alpha$ Survey (SHS; Parker et al. 2005, hereafter Paper I) provides high-resolution, narrowband, digital imaging data of the Southern Galactic Plane (SGP), making it a powerful tool for detecting  emission nebulae over a range of angular scales from arcseconds to degrees.  The survey utilised the 1.2-m UK Schmidt Telescope (UKST) at Siding Spring Observatory between 1998 and 2003.  Exposures were taken through a monolithic, high specification interference filter with a clear aperture of 305\,mm, a central wavelength of 6590\,\AA, a peak transmission of $\sim$90\%, and a full width at half maximum (FWHM) of 70\,\AA\ (Parker \& Bland-Hawthorn 1998). This  was sufficient to cover the H$\alpha$ and \NII\ emission lines across the full range of radial velocities observed in the Galaxy ($\pm$\,300\,km\,s$^{-1}$).  Each SHS field consisted of a 3-hr H$\alpha$ Tech-Pan exposure and a matching, usually contemporaneous, 15-min broadband (5900--6900\AA) short-red (SR)\footnote{The effective wavelength of the SR band is blueward of the widely used $R_F$ system (III-aF emulsion + RG630 filter;  Couch \& Newell 1980). The colour equations of this system are outlined by Morgan \& Parker (2005).} exposure taken through an OG590 filter for continuum subtraction purposes.  

The SHS footprint covers $\sim$4000 sq. degrees of the SGP between Galactic longitudes, $l$ = 210\arcdeg -- 30\arcdeg, and to a latitude, $|$b$|$ $\approx$\,10 -- 13\arcdeg.  It comprises 233 distinct but overlapping $5$\arcdeg\ survey exposures  on non-standard field centres $4$\arcdeg\ apart (cf. the normal 5\arcdeg\ spacing for UKST fields), together with 40 fields covering a separate contiguous region of 650 sq. degrees in and around the Magellanic Clouds (Morgan, Parker \& Phillipps 1999).   The scanned SHS data at 10\,$\mu$m (0.67\arcsec) resolution are available online in digital form through the SHS website\footnote{\url{http://www-wfau.roe.ac.uk/sss/halpha/}}, hosted by the Royal Observatory, Edinburgh.  To date the SHS is the only high-resolution \ha\ survey covering the full SGP.  Earlier work by J. Meaburn and co-workers used a square multi-element interference filter on the UKST, with a FWHM of 105\AA\ when used with the 098-04 emulsion (Elliott \& Meaburn 1976), but the plates only covered  limited areas of the SGP (e.g. Elliott, Goudis \& Meaburn 1976; Davies et al. 1978; Meaburn \& White 1982).  However, the coverage of the Magellanic Clouds was relatively complete (Davies, Elliott \& Meaburn 1976; Meaburn 1980), but these plates are not generally available.

The magnitude limit of the SHS is $R_S\simeq$\,20.5 for continuum point sources, and has a faint sensitivity limit to diffuse emission of 2--5\,rayleighs\footnote{The rayleigh (R) is a unit of photon flux (Hunten, Roach \& Chamberlain 1956; Baker 1974; van Tassel \& Paulsen 1975), where 1\,R = 10$^{6}$/4$\pi$\,photon\,cm$^{-2}$\,s$^{-1}$\,sr$^{-1}$ = 2.41\,$\times$ 10$^{-7}$\,erg\,cm$^{-2}$\,s$^{-1}$\,sr$^{-1}$ at H$\alpha$.}  (Paper~I).  This can be compared to the limit of the best Palomar Observatory Sky Survey (POSS) red plates, on the 103a-E emulsion, of $\sim$20\,R (see Peimbert, Rayo \& Torres-Peimbert 1975).  Both the POSS-II  (Reid et al. 1991) and UKST red plates used the IIIa-F emulsion; these plates have a sensitivity limit of $\sim$10--12\,R.  In comparison, sky-limited narrowband CCD limits are at the $\sim$1\,R level (e.g. Xilouris et al. 1996), comparable to the Southern H$\alpha$ Sky Survey Atlas (SHASSA; Gaustad et al. 2001) and the Virginia-Tech Spectral-line Survey (VTSS; Dennison, Simonetti \& Topasna 1998).   Note that the SHS coverage is over a factor of two larger than the equivalent northern INT Photometric \ha\ Survey (IPHAS; Drew et al. 2005), which only extends to $|$b$|$ $\sim5$\arcdeg.  The new VST/OmegaCam Photometric \ha\ Survey of the Southern Galactic Plane (VPHAS+; Drew et al. 2014)\footnote{\url{http://www.vphasplus.org/}}  is the southern counterpart of IPHAS, and also only extends to $|$b$|$ $\sim5$\arcdeg.  VPHAS+ has been underway on the ESO VST telescope in Chile since the start of 2012,  but will only supersede the SHS in the Galactic bulge and inner plane ($|$b$|$ \lessim\ 5\arcdeg) regions.   For reviews of current and upcoming \ha\ surveys, see Paper~I and Frew, Boji{\v c}i{\'c} \& Parker (2013, hereafter FBP13). 

The scientific exploitation of the SHS is still underway, as it has proven to be an excellent discovery medium for both extended nebulae and compact \ha\ emitters.  One area of success has been in  the detection of unprecedented numbers of new Galactic planetary nebulae (PNe), published in the two Macquarie/AAO/Strasbourg H$\alpha$ (MASH) catalogues (Parker et al. 2006; Miszalski et al. 2008).  While large numbers of these PNe are concentrated in the Galactic bulge (Acker, Peyaud \& Parker 2006), many are found throughout the survey footprint along the SGP.   These discoveries are having a major impact on PN science (see Parker et al. 2012a, for a review), and when combined with recent IPHAS discoveries (e.g. Mampaso et al. 2006; Viironen et al. 2009, 2011; Sabin et al. 2010; Corradi et al. 2011), and other significant contributions (e.g. Jacoby et al. 2010; Kronberger et al. 2012), have pushed the currently known Galactic population up to more than 3200 PNe (FBP13), more than double what it was a decade ago.   The SHS  also proved useful for finding previously uncatalogued emission features such as knots, jets and outer halos around known PNe (e.g. Miszalski et al. 2009; Frew, Boji{\v c}i{\'c} \& Parker 2012) and symbiotic stars (e.g. Miszalski et al. 2012), and for differentiating PNe from their mimics (Frew \& Parker 2010; Frew et al. 2010; De Marco et al. 2013).  

The survey has likewise been a boon for uncovering the optical counterparts of known Galactic supernova remnants (SNRs) seen previously only in the radio (Stupar et al. 2007a,b;  Stupar \& Parker 2009, 2011, 2012; Robbins et al. 2012).  Searches by Stupar, Parker \& Filipovi{\'c} (2008) also revealed many faint SNRs and candidate SNRs not previously recognised in the radio (see also Walker, Zealey \& Parker 2001; Alikakos et al. 2012).  Similarly, the survey's potential for optical detections of SNRs in the LMC has been realised by Boji\v{c}i\'c et al. (2007) and Reid et al. (in preparation).  
The SHS has also been used to study star-forming regions (e.g. Cohen et al. 2002, 2007, 2011; Zavagno et al. 2007; Urquhart et al. 2007; Cappa et al. 2008; Deharveng et al. 2009; Pinheiro et al. 2010; Russeil et al. 2010; Bik et al. 2010; Murphy et al. 2010), the diffuse interstellar medium (ISM; Tackenberg et al. 2013), Herbig-Haro objects and related outflows (Girart \& Viti 2007;  Mader et al. 1999), and circumstellar nebulae around massive stars (e.g. Cohen, Parker \& Green 2005; Gvaramadze, Kniazev \& Fabrika 2010; Stock \& Barlow 2010; Duronea, Arnal \& Bronfman 2013).  
Finally, the SHS has  revealed a range of emission-line stars including WR stars (Drew et al. 2004; Hopewell et al. 2005; Miszalski, Miko\l ajewska \& Udalski 2013), Be and T Tauri stars (Pierce 2005), cataclysmic variables (Pretorius \& Knigge 2008), and symbiotic stars (Boissay et al. 2012; Miszalski et al. 2013).  Importantly, using the  related SHS imaging data for the central region of the LMC, Reid \& Parker (2006a, b) uncovered over 2,000 compact emission sources, including 450 new PNe and many emission-line and variable stars (Reid \& Parker 2012).   A new star-forming ring galaxy 
has also been found from SHS images (Parker et al. 2013).

Despite these studies, the full potential of the SHS has been hampered by the lack of a clear procedure to establish a reliable, integrated H$\alpha$ flux or a surface brightness for the any distinct emission structure visible on the survey.   Thus we need a reliable method to transform the pixel intensity values from SuperCOSMOS microdensitometer scans (Hambly et al. 1998) of the \ha\ images into intensity units, which should also be consistent from field to field.  Pierce (2005) and Paper\,I showed that the survey data have been well exposed to capture Galactic emission on the linear part of the photographic characteristic curve (e.g. Smith \& Hoag 1979; Parker \& Malin 1999).   However, no detailed process for achieving reliable flux calibration for discrete sources on the SHS has yet been provided in the literature. 
To address this deficiency we extend this earlier work so researchers can derive meaningful results from flux-calibrated SHS data as required, following the precepts outlined in this paper.  The motivation for this study is to explore the viability of using the SHS imaging data to establish a reliable mechanism for \ha\ flux measurements.   
This study also attempts to address the various limitations of the base data, including the restricted dynamic range (when scanned by the SuperCOSMOS microdensitometer) compared to the inherent density range of the original photographic films (e.g. MacGillivray \& Stobie 1984; Hambly et al. 2001).   The structure of this paper is as follows:   \S2 describes the limitations of the photographic survey data and \S3 describes the flux calibration process for the SHS.  In \S4 we focus on the derivation of SHS \ha\ fluxes from the imaging data, and in \S5 we undertake a comparison of the SHS fluxes with independent measurements.  Our conclusions and recommendations are given in \S6.


\section{Flux Calibration of the SHS survey images}
The calibration of the SHS survey images is achieved by direct  comparison of the continuum-subtracted SHS data with the matching continuum-subtracted SHASSA data (Gaustad et al. 2001) which provides narrowband (actually H$\alpha$ + \NII) and continuum images of the entire southern sky.   The preliminary flux calibration process was described by Pierce (2005) and Paper~I, and the reader is referred to the latter paper for technical details.   Although SHASSA has a low spatial resolution with 48\arcsec\ pixels, it has the considerable advantage of being continuum subtracted and accurately flux calibrated, using a range of calibrating PNe from Dopita \& Hua (1997).   The zero-point calibration of SHASSA as defined by Gaustad et al. (2001) and is good to better than 9 per cent.  This was confirmed using the independent data from the Wisconsin H-Alpha Mapper (WHAM; Haffner et al. 2003) Northern Sky Survey\footnote{\url{http://www.astro.wisc.edu/wham/}} at 1\arcdeg\ resolution.  The SHASSA zero point of SHASSA was independently confirmed by Finkbeiner (2003), also using WHAM data.  However, this comparison could only be done north of $\delta$ = $-30$arcdeg, the southern limit of the WHAM survey.  Below this declination, the SHASSA flux calibration was extended using a cosecant law fit (Gaustad et al. 2001), and so may not be particularly reliable on the largest angular scales (\grtsim5\arcdeg).  The WHAM observations are being extended to the southern sky as the WHAM-South Survey (Haffner et al. 2010), which will allow an accurate calibration of the far southern sky when it is eventually completed. 
Pierce et al. (2004), Pierce (2005), Frew (2008), and FBP13  also provide independent analyses of the SHASSA data.  Indeed this work has recently culminated in the use of the SHASSA (and complimentary VTSS) data to provide accurate, integrated H$\alpha$ fluxes for over 1250 Galactic PNe (see FBP13).   

\begin{figure*}
\begin{center}
\includegraphics[width=17.7cm]{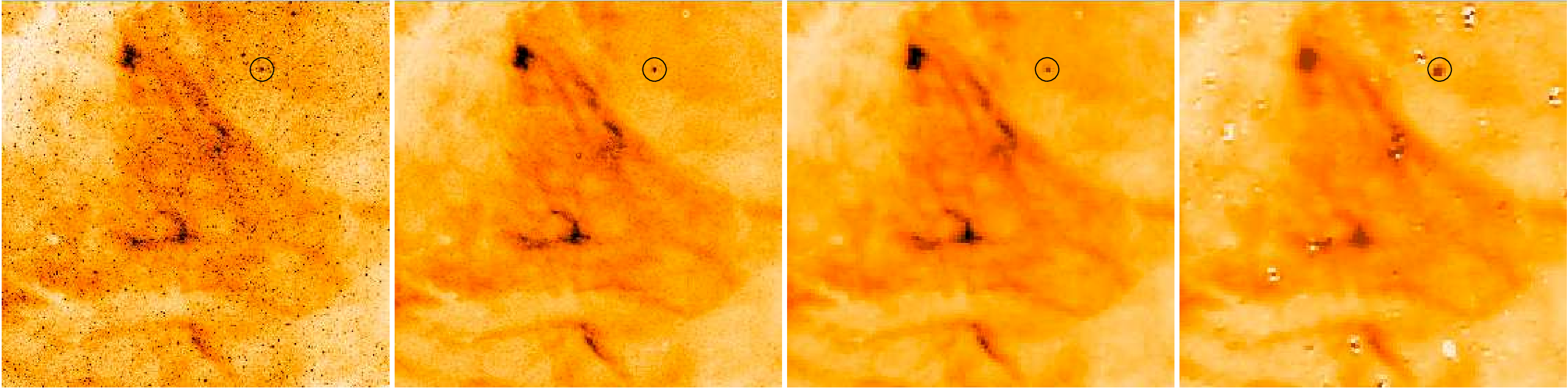}
\caption{Comparison of SHS and SHASSA data of a 80\arcmin\ $\times$ 80\arcmin\ region near the centre of SHS field HAL0970 at RA,DEC (J2000) 18$^{\rm h}$29$^{\rm m}$, $-16$\arcdeg00\arcmin.  This field is identified as a dashed box labelled `a' in Figure\,2.  From left to right  (on the same relative scale):  (1)  \ha\ image; (2) matching continuum-subtracted \ha\ image; (3) rebinned continuum-subtracted SHS \ha\ image; (4) equivalent flux-calibrated SHASSA \ha\ image.  Note the stellar residuals in the SHASSA image, and the planetary nebula M\,1-46, highlighted with a circle in each panel. }
\label{fig:SHS_vs_SHASSA}
\end{center}
\end{figure*}

In this paper we have completely redone the SHS calibration process afresh, by selecting data from the unvignetted central regions (see Figure~\ref{fig:HAL0970}) of all 233 SHS fields (cf. Paper\,I).  This was undertaken using the online\footnote{\url{http://www-wfau.roe.ac.uk/sss/halpha/hafields_online.html}} 16$\times$ blocked-down data for each field, which have a resolution of 10.7\arcsec\ pix$^{-1}$.  These images  provide a  convenient way of working across large areas for comparison with the calibrated SHASSA data, but were not available in that format to Pierce (2005).  It is important to remember  that one 13\arcdeg\,$\times$\,13\arcdeg\  SHASSA field completely covers a single 5\arcdeg\,$\times$\,5\arcdeg\ SHS field, and that the SHASSA data suffer from artefacts resulting from the effects of bright stars (see Gaustad et al. 2001; and Figure~\ref{fig:SHS_vs_SHASSA}).

\subsection{Continuum Subtraction on SHS Images}
Each matching SR image provides an effective measure of the continuum component for each SHS field, despite including the \ha\ (and the adjacent \NII) emission lines in the bandpass.   The issue of \NII\ contamination and how it can be accounted for is described in \S\,\ref{sec:deconvolve}.  
The \ha\ and SR images are generally exposed to attain the same depth for continuum point sources (Paper~I).  However, the nature of photography and the vagaries of the observing conditions (e.g. if the exposure pairs were not contemporaneous and taken in different moon phases or if the seeing changed) mean that the SR and \ha\ point source depth image quality is similar, but not exactly so.  Thus, it is necessary to precisely determine a continuum-subtraction scaling factor (SF) on a field-by-field basis for the continuum subtraction process to be effective.  For high-dynamic range CCD exposures, the standard method for determining the appropriate scaling factor to subtract continuum from narrow-band is to compare aperture photometry for stars on both images whose exposures are normally interleaved on short time-scales.  This does not work so well with the \ha\ plus SR film exposure pairs (Pierce 2005), often leading to under- or over-subtraction of the continuum.  This produces positive or negative residuals around point sources, so care is needed (see \S\ref{sec:measurement_pipeline}).

Firstly, the 16$\times$ blocked-down (10.7\arcsec\,pix$^{-1}$)  H$\alpha$ and SR survey images for each SHS field were obtained from the SHS website.
 Then, a continuum-subtracted \ha\ (+ \NII) image was created from the matching \ha\ and SR images in {\tt IRAF}\footnote{IRAF is distributed by the National Optical Astronomy Observatory, which is operated by the Association of Universities for Research in Astronomy, under cooperative agreement with the National Science Foundation.}, using a scaling factor (SF), as defined in Pierce (2005) and Paper~I.   A range of values for the scaling factor can be applied to the SHS data until the zero-point of the continuum-subtracted SHS survey images best match the zero-point of the equivalent SHASSA data, indicating the appropriate value to use.  The accuracy of the scaling factor is critical in order to avoid incorrect subtraction of the continuum component.  
 Once a suitable SF has been determined, the continuum subtraction process works well, removing most of the stellar images and the diffuse continuum.  Only stars that sit in regions with the strongest \ha\ emission have been over-subtracted, and hence appear as white spots in the image, as do any high-surface brightness knots  in fainter extended nebulae.  For a few fields, we adopt SF = 1, if the comparison is inadequate to determine a formal value.  

For example, Figure~\ref{fig:SHS_vs_SHASSA} shows four images of a 80\arcmin\ square region from field HAL0970, which shows \ha\ emission at a range of intensities. The left image is the 16$\times$ blocked-down SHS \ha\ image, the second image is the continuum-subtracted \ha\ image, the third image shows this image re-binned to a resolution to match SHASSA, while the right image shows the equivalent SHASSA data. In this example, this continuum subtraction process has worked well, removing most of the stellar images and the diffuse continuum. Only the stars that sit in the strongest emission have been over-subtracted and appear as white spots in the image.    Note the residuals around stars seen in the SHASSA image, and the planetary nebula M~1-46, marked with a circle.
\subsection{Calibration Factors}\label{sec:cal_fac}
Each SHS field has a corresponding calibration factor (CF) determined by Pierce (2005), but these are not generally available to the community.  Consequently, we have first re-calculated and improved the reliability of the CF for each survey field using our somewhat different approach based on the 16$\times$ binned SHS data.  In addition, in fields lacking obvious large scale emission, we have attempted to identify isolated nebulae (such as PNe) in order to provide a new calibration.  

For each SHS field, we selected the corresponding SHASSA field by minimising the distance between the respective field centres.    Secondly, the 16$\times$ blocked-down continuum-subtracted images were re-binned and re-aligned  to match the corresponding 48\arcsec\-resolution SHASSA image using the {\tt MONTAGE} toolkit (http://montage.ipac.caltech.edu/).  The SHASSA image was then cropped to the re-projected and re-binned SHS image.  Now, the pixel intensity counts of the re-binned SHS images can be compared directly with the equivalent pixel intensity values (in R) from SHASSA to yield the CF for that field.    The {\tt MONTAGE mProject} module also returns the average value of binned pixels, which means that the final CF, derived from the pixel-to-pixel comparison with SHASSA, applies to the original full resolution SHS image.    

\begin{figure}
\begin{center}
\includegraphics[width=9.25cm]{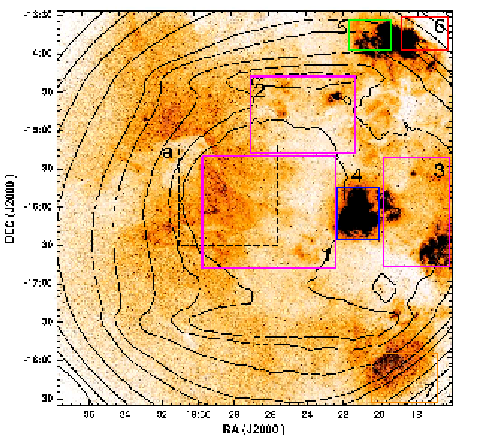}
\caption{The full 5\arcdeg$\times$5\arcdeg\ SHS blocked-down continuum-corrected \ha\ image of field HAL0970, overlaid with contours of the filter response at 70\%, 75\%, 80\%, 85\%, 90\%, 92\%, 95\%, 97\% and 98\% of peak.  Selected subregions are marked by rectangles as described in the text.  Those numbered 1--4 are taken from the best area of filter response and were chosen to select a range of emission line intensities. Regions 5--7 probe the edges of the field where the filter transmittance drops below 85\% and are not normally accessed by the online tool.  The dashed box labelled 'a' near the field centre is the for the region shown in Figure\,1.   A colour version of this figure is available in the online journal.} 
\label{fig:HAL0970}
\end{center}
\end{figure}

Large (typically 30\arcmin\ -- 60\arcmin) representative areas from the normal unvignetted area of each SHS field were selected and matched to give a robust calibration for each field.  We emphasise that these regions were generally available for calibration, due to the large field-to-field overlap of the survey.  This known vignetting from the \ha\ filter was the original motivation for the SHS field centres being placed on a 4\arcdeg\ grid rather than the standard 5\arcdeg\ grid for Schmidt plates, so this problem can be avoided (see Paper~I).  
Figure~\ref{fig:HAL0970} shows the blocked-down SHS \ha\ image of field HAL0970 with overlaid contours of filter response at 70\%, 75\%, 80\%, 85\%, 90\%, 92\%, 95\%, 97\% and 98\% of peak transmission.  Selected subregions are marked by rectangles of different colour.   The magenta rectangles represent three subregions (numbered 1--3) inside the unvigneted part of the field, and the blue rectangle (region 4) includes the bright \HII\ region M~17 (NGC~6618), also located within the unvignetted part of the field.   For a comparison, the green, red and orange rectangles (regions 5--7) are areas of the field affected by filter vignetting.

A graphical plot of continuum-subtracted SHS pixel intensities against equivalent SHASSA pixel intensities should return a linear relation with a common zero point if the reduction and intensity calibration have been properly carried out and assuming the photographic intensity values are unsaturated and fall on the straight-line portion of the characteristic curve.  The gradient of this plot is the CF for the image (see below).  
If the CF determined from the comparison plot is well behaved, it is applicable to full resolution SHS data for the entire field.  Unfortunately, some survey fields exhibited very little diffuse emission making calibration for these fields less straightforward.  In such case a careful search for small areas of discrete, low intensity \ha\ emission in the unvignetted zone of the field was made, including any known large planetary nebulae.  This enabled pixel by pixel comparisons between the two surveys to deliver appropriate CFs.

\begin{figure}
\begin{center}
\includegraphics[width=9.15cm]{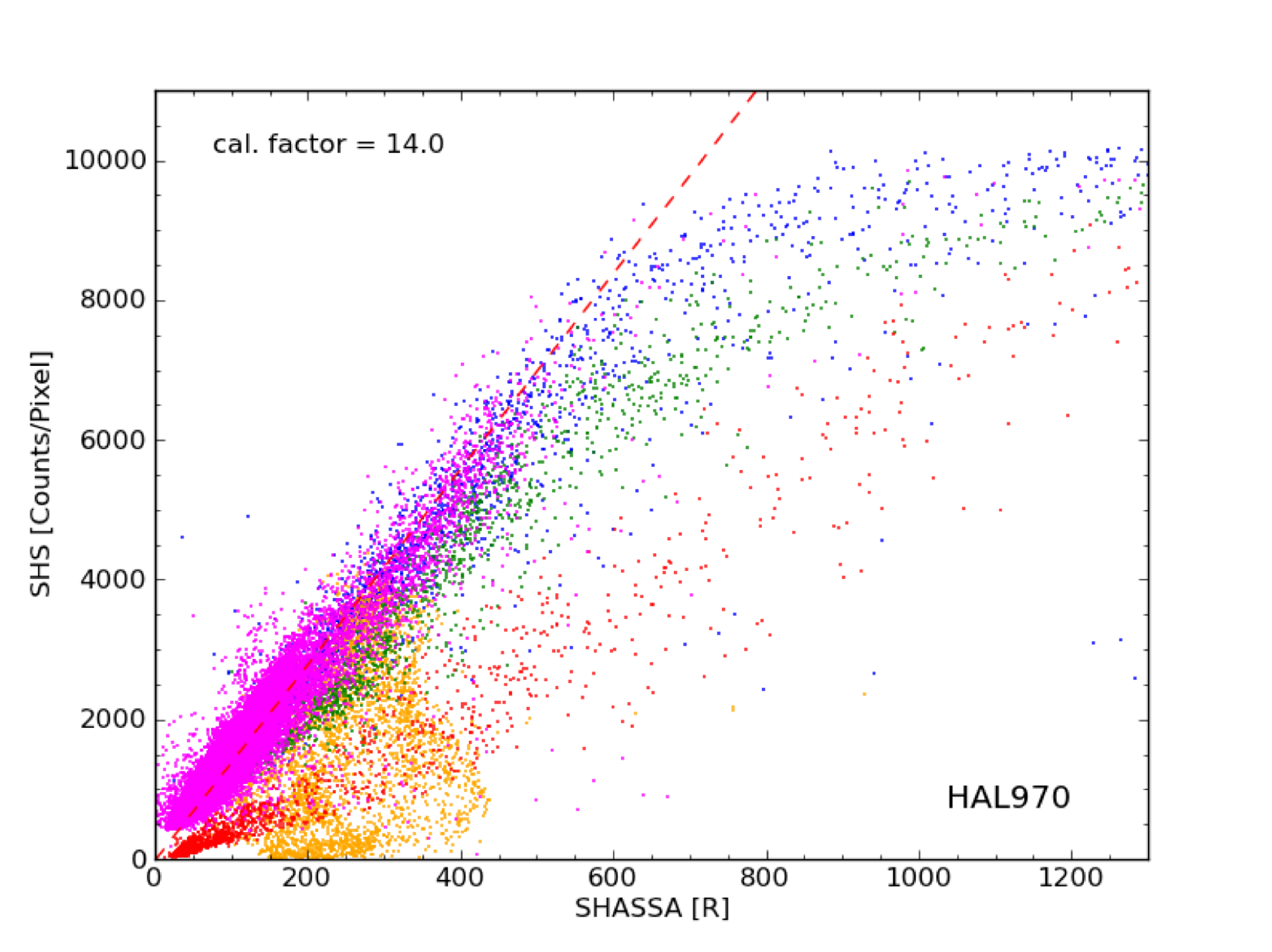}
\caption{\small Calibration plot of the numbered and colour-coded subregions on field HAL0970.  The magenta and blue points are from four subregions (1--4) inside the unvignetted part of the field, with the blue points representing the bright \HII\ region M~17.  These align with the magenta points, but show the effect of bright photographic saturation, above 500\,R.  The green, red and orange points show data from areas of the field (regions 5--7) affected by vignetting, and these are displaced from the magenta/blue points.  A linear fit to the unvignetted data (between 50 and 500\,R) is also plotted (dashed line), with a gradient (CF) of 14.0\,counts\,pix$^{-1}$\,R$^{-1}$.  
} 
\label{fig:HAL0970_plot}
\end{center}
\end{figure}

The process is illustrated for field HAL0970 (Figure~\ref{fig:HAL0970}).  We show the calibration plot of the selected rectangular subregions as Figure~\ref{fig:HAL0970_plot}.  The points from each subregion are plotted as a different colour.   The magenta points are from three subregions (1--3) inside the unvignetted part of the field, and the blue points represent the bright \HII\ region M~17 (region 4), also located within the unvignetted part of the field.  These align with the magenta points on the one trend, but show the effect of bright photographic saturation, seen above 500\,R.  A least-squares linear fit to the magenta and blue data points below 500\,R is also plotted, with a gradient of 14.0\,counts\,pix$^{-1}$\,R$^{-1}$.   This demonstrates that if provided areas are chosen from within the unvignetted field, consistent CFs are obtained giving confidence in its applicability across the field. However, if pixel-pixel comparisons are made from the sub-regions in Figure~\ref{fig:HAL0970_plot}, which sample data from the vignetted regions of the field (green, red and orange points; regions 5--7), the agreement is poor and varied.  The spread in counts is artificial and results from applying the flat field correction to the underexposed areas of film.

\begin{figure}
\begin{center}
\includegraphics[width=9.15cm]{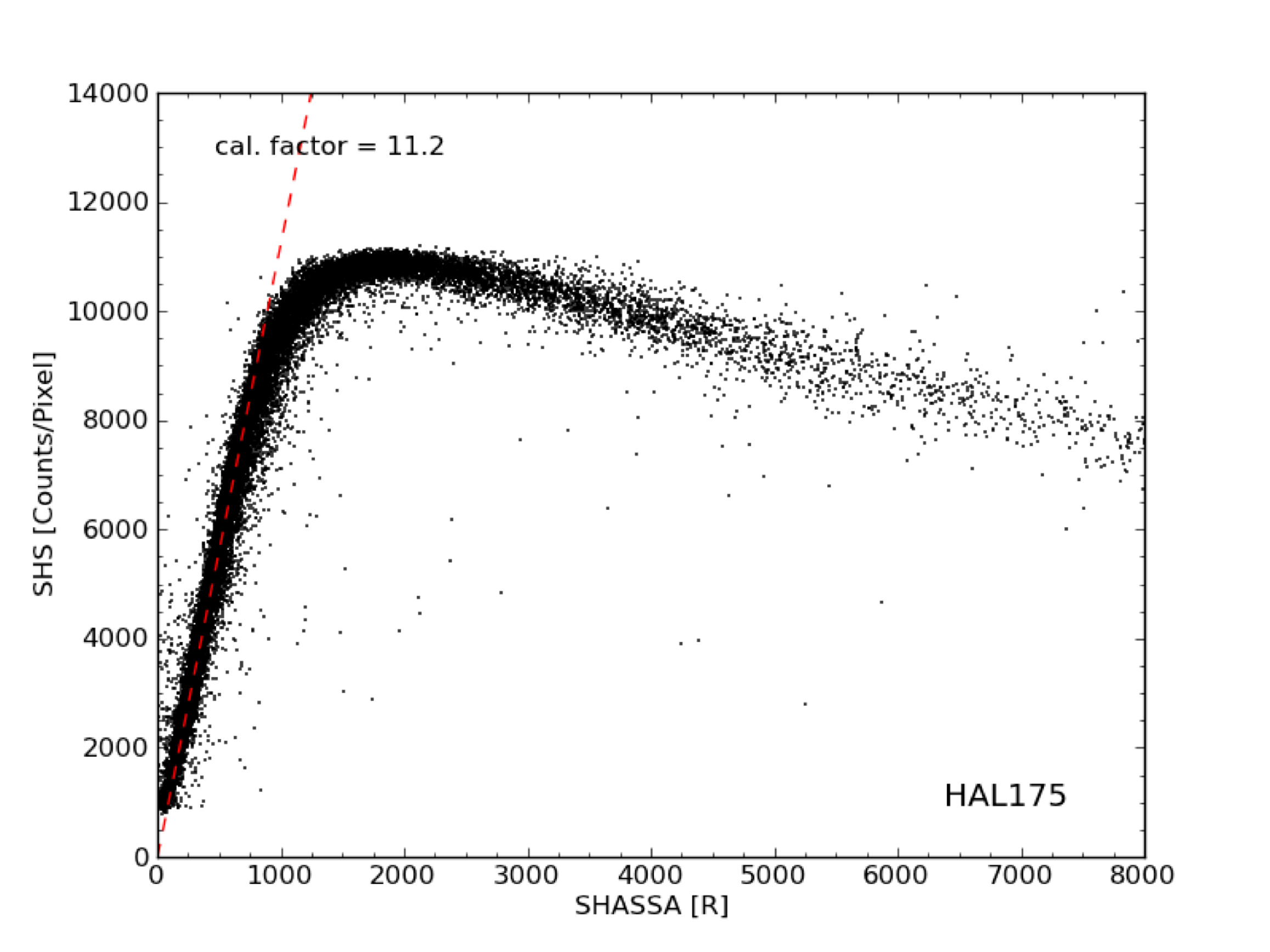}
\caption{\small Calibration plot of field HAL0175, which is a field of intense \ha\ emission centred on the Carina Nebula (NGC 3372).  Owing to this the \ha\ exposure was reduced to two hours.  The plot illustrates beautifully the effect of flux over-subtraction at high surface brightness.  A linear fit to the data between 50 and 700\,R is also plotted, with a gradient (CF) of 11.2\,counts\,pix$^{-1}$\,R$^{-1}$.  A colour version of this figure is available in the online journal.}
\label{fig:HAL0175_plot}
\end{center}
\end{figure}

We also illustrate the calibration plot for HAL0175, which is a field of intense \ha\ emission centred on the Carina nebula (Figure~\ref{fig:HAL0175_plot}).  Because of this, the \ha\ exposure was reduced to two hours, so  photographic saturation sets in at a higher level of 700 -- 800\,R.  The plot illustrates beautifully the effect of flux over-subtraction at high surface brightness, as the SR `continuum' image also saturates in the brightest section of the Carina nebula.   A linear fit to the data between 50 and 700\,R is plotted as a red dashed line, with a gradient (CF) of 11.2\,counts\,pix$^{-1}$\,R$^{-1}$. 
This calibration process was repeated for all 233 SHS fields.  As a result, more fields now have reliable CFs than were determined by Pierce (2005), and the data are presented in Table~\ref{tab:main} after being assigned individual quality flags on a ranking scale of 1, 2, or 3. 

First-ranked fields have a reliable CF, that, in most cases is derived from SHS pixel data that exhibits a completely linear relation with equivalent SHASSA data over much of the data range. Such fields are considered to produce the most reliable results.  Fields ranked two indicate non-linear pixel relations between the SHS and SHASSA and/or are from SHS fields that exhibit relatively poor dynamic range or suffer from photographic fog.  These fields nevertheless do have a reliable CF, and should still produce a good calibration. Rank three is assigned to fields without a reliably estimated CF and display one or more of the following anomalies: early saturation at the bright (large intensity) end of the data's dynamic range, small dynamic range, either due to the absence of diffuse \ha\ emission, or a consequence of elevated levels of photographic fog. For the rank three fields we applied a default value for the SF of 12.0$\pm$3.0 counts~pix$^{-1}$~R$^{-1}$.

\begin{figure*}
\begin{center}
\includegraphics[width=17.25cm]{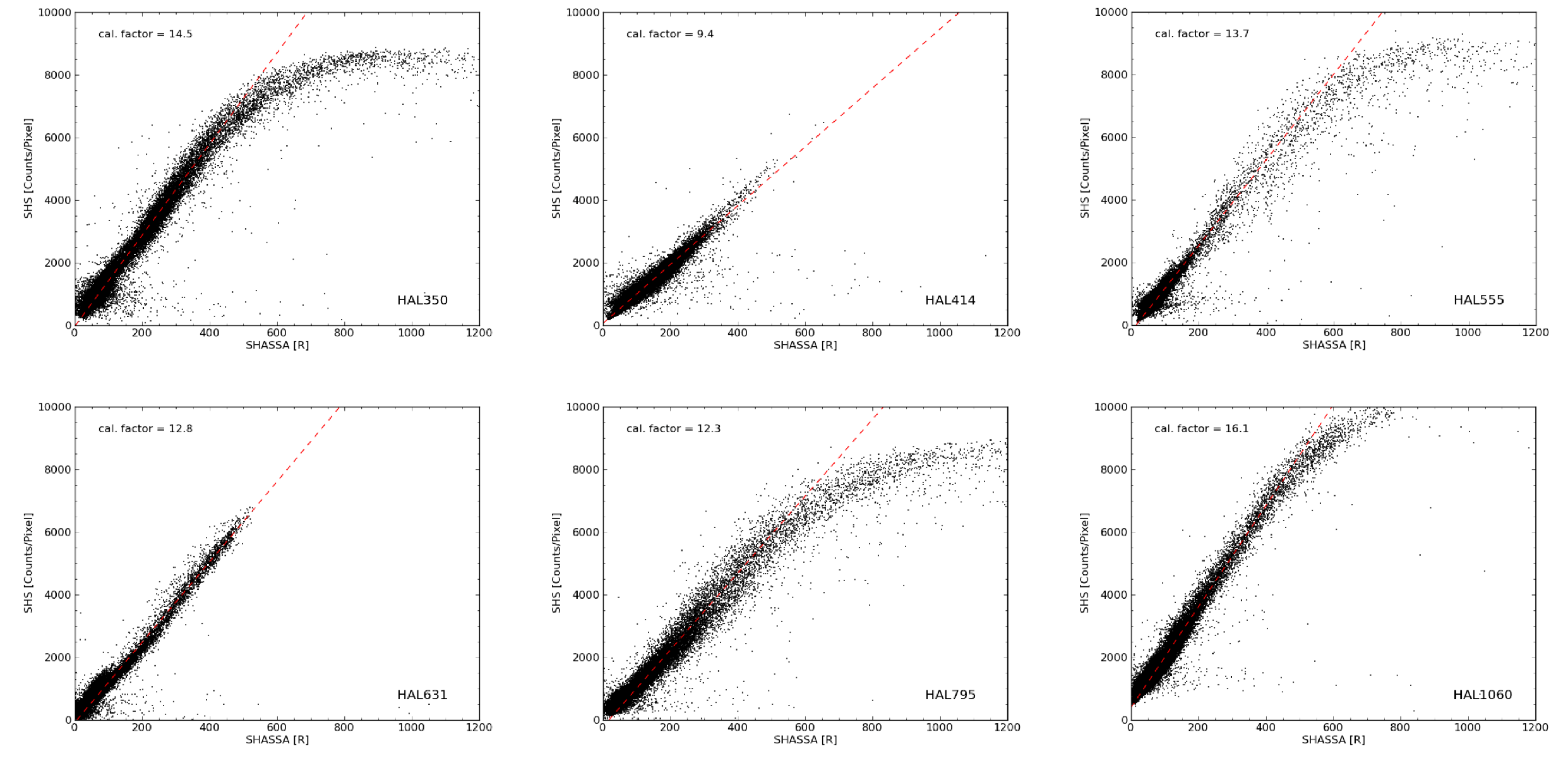}
\caption{SHASSA-SHS pixel-by-pixel comparison plots for six first-ranked SHS fields.    Some SHS fields (e.g. HAL0350, HAL0482 and HAL0555) show evidence of saturation where the intensity exceeds $\sim$500--600\,R.   Below this level, most fields show a linear relation between the SHASSA and SHS.   Note that fields HAL0134 and HAL1060 show evidence of photographic fog, which elevates the sky level on these SHS fields.}
\label{fig:comparison_plots}
\end{center}
\end{figure*}

\begin{figure*}
\begin{center}
\includegraphics[width=17.25cm]{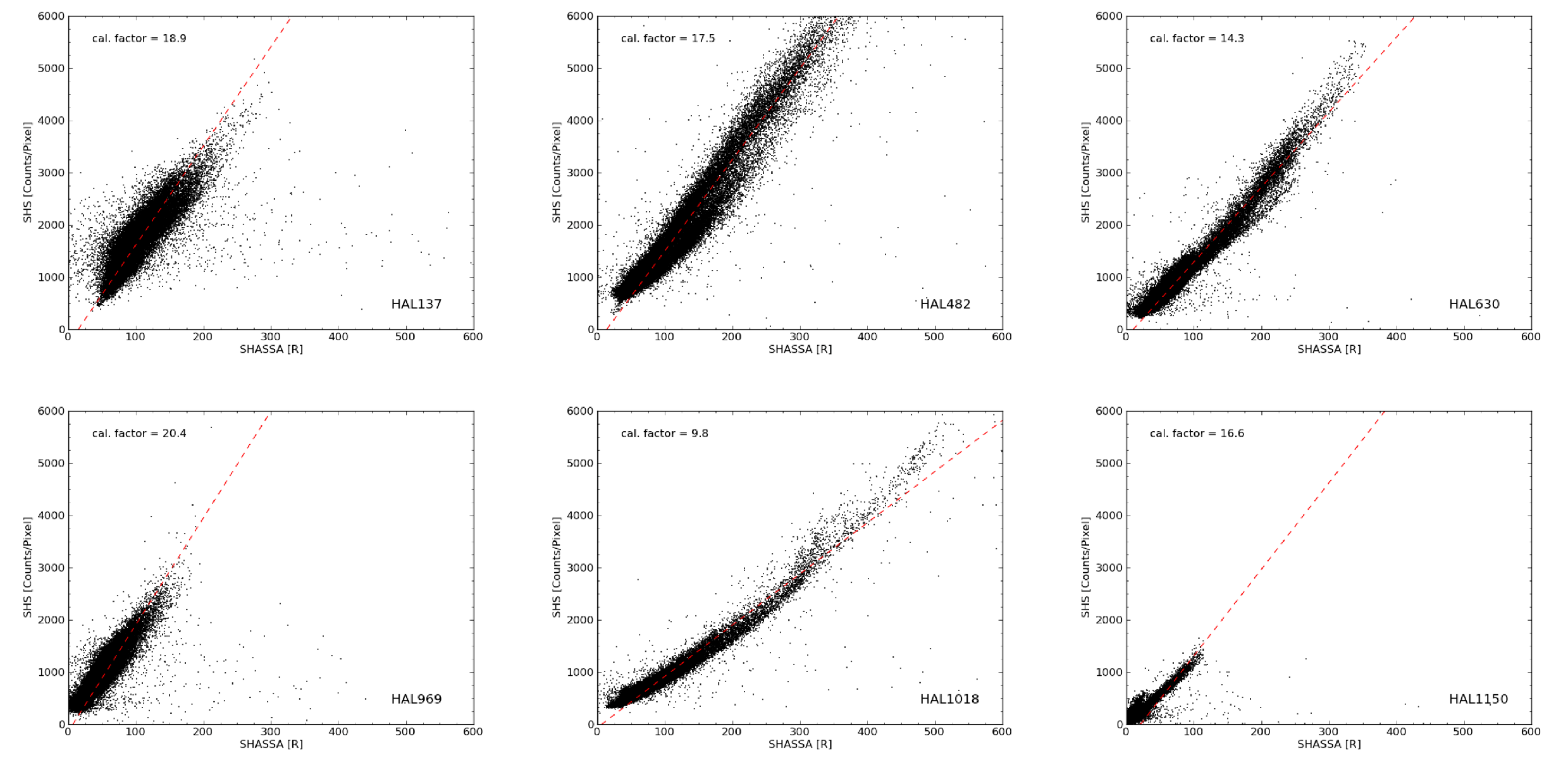}
\caption{SHASSA-SHS pixel-by-pixel comparison plots for six second-ranked SHS fields.  The slight inconsistencies around the linear fit seen in HAL0137 are due to the slightly varying filter response in different areas of the field where the discrete \HII\ regions occur. HAL1150 is an example of a field with a limited intensity range of emission, while the other five fields show varying levels of photographic fog.  }
\label{fig:comparison_plots2}
\end{center}
\end{figure*}

The comparison plots for six well-behaved first rank SHS fields are shown in Figure~\ref{fig:comparison_plots}.  Here an essentially linear pixel-intensity relation can be seen between the SHS and SHASSA data and the linear fits are overlaid.   A further six fields with some caveats are illustrated in Figure~\ref{fig:comparison_plots2}.  In Figure~\ref{fig:comparison_plots2}, we  show the comparison plots for six second-ranked SHS fields, which have linear fits that are not as precise as the first-ranked fields.    In the case of HAL1018, subtle changes in the \NII/\ha\ ratio across an ionization front in the large \HII\ region IC~2177 will produce a variable ratio of SHS counts R$^{-1}$ across the nebula.   Furthermore, in other fields, discrete but widely separated \HII\ regions may occur in areas of the filter response that vary at the 2--3\% level.  HAL0137 is such a case, where the slight inconsistencies around the linear fit seen are due to the slightly varying filter response in different areas of the field where the discrete \HII\ regions occur.   HAL1150 is an example of a field with a limited intensity range of emission, while the other fields show varying levels of photographic fog (Tritton 1983), which restricts the dynamic range of these survey exposures. 

These various effects can conspire to produce a formal fit with apparent curvature, or the appearance of two or more closely spaced trends on the calibration plot.  These effects are usually small however, and should not compromise the utility of the fit in these SHS fields.  The CFs determined from the linear fits can be applied with some confidence to the full (0.67\arcsec) resolution pixel data for the relevant survey field.

\begin{table*}
\begin{center}
\caption{Data summary for all 233 SHS survey fields.   Note there are 63 fields with no formal CF determined (usually due to poor dynamic range or high chemical fog, or both). A default value of 12.0 $\pm$ 3.0 counts\,pix$^{-1}$\,R$^{-1}$ can be used for these fields, as explained in the text.  The table is published in its entirety as an online supplement, and a portion is shown here for guidance on its form and content.} 
\label{tab:main}
\begin{tabular}{ccccl}
\hline 
Field 		&  ~~~~~SF~~~~~ 	&  CF 		&  ~~Quality~~  	& Comment   \\
  			&   		&(count\,pix$^{-1}$R$^{-1}$)   &     		&   \\
\hline \noalign{\smallskip}
HAL0067 	&	...	&     (12.0)	&	3	&	small dynamic range	\\
HAL0068 	&	...	&	(12.0)	&	3	&	small dynamic range	\\
HAL0069 	&	...	&	(12.0)	&	3	&	small dynamic range	\\
HAL0070 	&	...	&	(12.0)	&	3	&	 photographic fog	\\
HAL0096 	&	...	&	13.3   	&	3	&	small dynamic range	\\
HAL0097	&	0.98	&	11.4	&	2	&	$\ldots$	\\
HAL0098 	&	0.83	&	13.9	&	2	&	 photographic fog	\\
HAL0099	&	0.42	&	(12.0)	&	3	&	 photographic fog	\\
HAL0100	&	1.10	&	12.7	&	1	&	$\ldots$	\\
HAL0101	&	1.12	&	11.7	&	2	&	small dynamic range	\\
$\vdots$       &	$\vdots$  &	$\vdots$  		&	$\vdots$  		&	~~~~~~~~~~$\vdots$  \\
\hline 
\end{tabular}
\end{center}
\end{table*}

\begin{figure*}
\begin{center}
\includegraphics[width = 16cm]{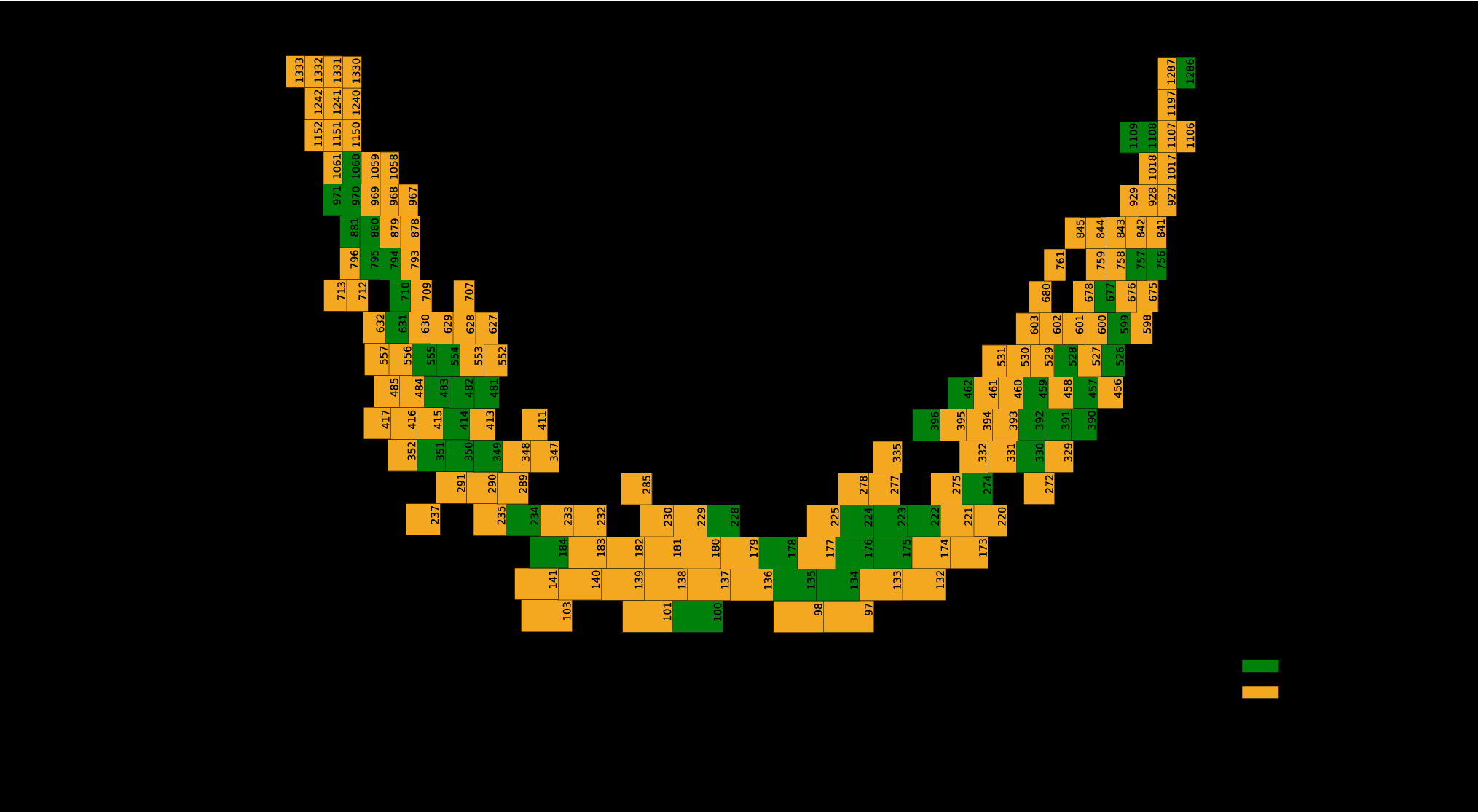}
\caption{The sky coverage of the SHS in equatorial J2000 coordinates (note that truncated RA and DEC coordinates).  The green, amber, and white boxes represent first-, second-, and third-ranked fields, respectively, and the field numbers are given.  A `clickable' version (though without colour-coding)  is available on the SHS website.   A colour version of this figure is available in the online journal.  }
\label{fig:coverage}
\end{center}
\end{figure*}

\begin{figure}
\begin{center}
\includegraphics[width=8.4cm]{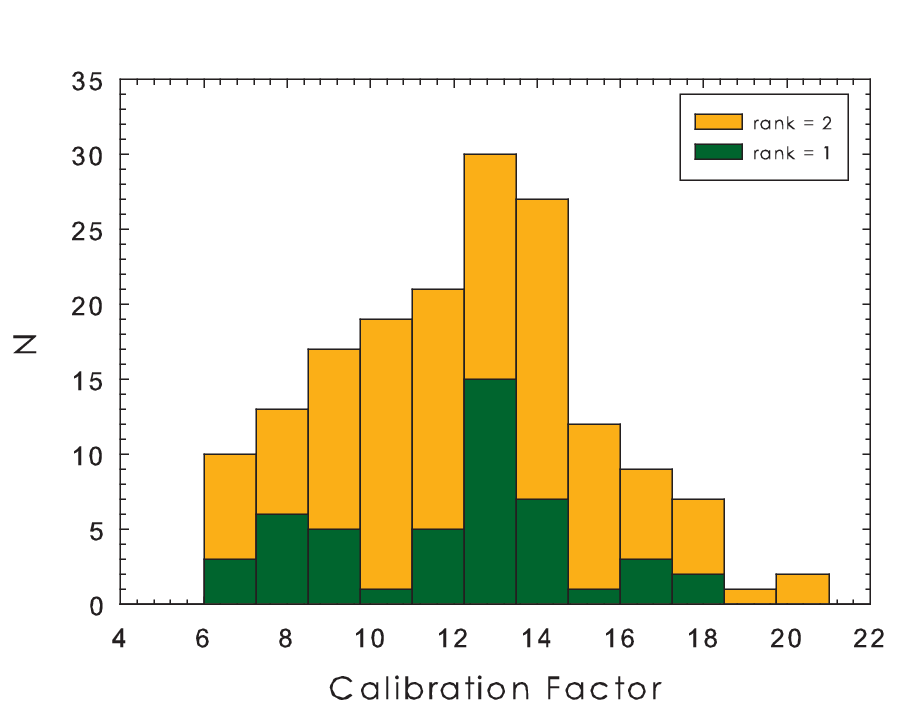}
\caption{Histogram of derived calibration factors (CFs) for 170 SHS fields.  The green and orange bars represent the numbers of first- and -second-ranked fields, respectively.  A colour version of this figure is available in the online journal.}
\label{fig:histogram}
\end{center}
\end{figure}

The full sky coverage of the SHS  is illustrated in Figure~\ref{fig:coverage}, where the quality flag for each field is colour-coded.  Note that the third-ranked fields are essentially devoid of diffuse \ha\ emission.  Figure~\ref{fig:histogram} shows the CF distribution for 170 fields (of 233) that could be fit with a linear function.  The distribution of the first ranked-fields has a mean value of 11.9 $\pm$ 2.9 counts\,pix$^{-1}$\,R$^{-1}$ (1$\sigma$ dispersion), while the second-ranked fields have a mean 12.2 $\pm$ 3.2 counts\,pix$^{-1}$\,R$^{-1}$.   The weighted mean is 12.0 $\pm$ 3.0 counts\,pix$^{-1}$\,R$^{-1}$, which can be used as a default value  for those fields without a formal CF determination.  This approach then provides a direct mechanism  to determine \ha\ fluxes for individual emission structures not resolved by SHASSA, including, for example,  many new compact PNe discovered from the SHS data (Parker et al. 2006; Miszalski et al. 2008).

\section{The Limitations of the SHS Survey Data}
As previously mentioned not all SHS fields provide reliable flux measurements  Some fields are also naturally devoid of extensive \ha\ emission, especially those at higher Galactic latitudes. In these cases it can be problematic to find unvignetted areas with emission features that demonstrate sufficient dynamic range  to provide a large enough lever-arm for a reliable calibration.  
Unlike CCD data, photographic exposures have a limited linear response and saturate at a photographic density\footnote{The photographic density, $D$, is the logarithm of the ratio of incident light to transmitted light through a region of a developed emulsion.} of $D\sim$\,4 (e.g. Hambly et al. 2001).  However, the automated SuperCOSMOS microdensitometer used to digitise the photographic exposures imposes its own limitations, restricting the density range even further to $D\leq$ 2  (Hambly et al. 1998).  The data scanning and calibration processes are affected not only by the relatively low dynamic range of SuperCOSMOS, but also because some H$\alpha$ and SR pairs are not contemporaneous, being taken on different nights, and under different seeing conditions and lunar phase.  Furthermore, these differences in exposure date may mean different hypersensitised emulsion batches were used, which may have different levels of chemical fog and speed.  Some fields have a higher than average level of photographic chemical fog due to the Tech Pan films being push-processed.  These factors can introduce variations in background and sensitivity from exposure to exposure within the same `matching' \ha\ and SR pair. 

Further uncertainties in the emission levels across survey fields can arise from the presence of continuum emission, night-sky auroral lines, and geocoronal \ha\ emission  as well as the effects of variable atmospheric extinction (Paper I).   Ideally, all of these components would need to be disentangled to extract the true Galactic \ha\ emission. In practice, the geocoronal and auroral emission is considered as a low-level,  temporally varying wash, which elevates the general background on each exposure to a slightly varying degree.   This is considered to be less than 2\,R across the SHS (Nossal et al. 2001), based on H$\alpha$ data from the WHAM Northern Sky Survey (Haffner et al. 2003).  

The original set of  pixel-by-pixel comparison plots for selected SHS fields given by Pierce (2005), demonstrated the spread of dynamic range and linearity exhibited by the SHS data on a field by field basis.   Many fields have a well defined linear trend but some are less well-behaved due to some of the factors previously mentioned. As a consequence fields were allocated a quality flag to reflect this diversity (see \S\,\ref{sec:cal_fac}).  This is based on a fresh examination of all survey fields where each SHS field was ranked according to suitability for obtaining accurate H$\alpha$ fluxes. 
These flags are given in Table~\ref{tab:main}.   Representative samples of first- and second-ranked fields are shown in Figure~\ref{fig:comparison_plots} and Figure~\ref{fig:comparison_plots2}.

In summary,  50 SHS fields (21\% of the total) are well constrained by a linear fit below 500\,R and so have the top ranking of 1, while a further 120 fields (52\%) also have a linear fit, but with some caveats, and have been given a  rank of 2.   The 63 remaining SHS fields (27\%) have little diffuse emission, making the determination of reliable CFs less straightforward, primarily due to a very restricted dynamical range of emission.  A default value of 12.0 counts\,pix$^{-1}$\,R$^{-1}$ can be safely used for these fields.  For ease of use, we give the CFs for 170 SHS fields in Table~\ref{tab:main}.  

\section{Reliability of the Flux Calibration}
In the previous section we have established revised calibration factors for each SHS  field.  In order to  demonstrate the capability of these fields to provide accurate integrated fluxes for discrete sources such as PNe,  \HII\ regions, WR shells, and Herbig-Haro objects, we identified a variety of suitable calibration objects across a range of SHS fields that have independent integrated H$\alpha$ fluxes from the literature, or determined by us from SHASSA measurements.  This process is described below.

\subsection{Selection of Individual Sources}
The object sample chosen for this flux calibration study includes a variety of astrophysical sources exhibiting a range of angular scales and emission intensities across the full range of SHS field exposure characteristics.  These should be sufficient to determine if the calibration factors derived in the previous section are correct.  The selected objects  range in angular size from a few arcseconds to nearly 2\arcdeg\ across, and cover a \ha\ flux range of over $\sim$7\,dex.  
The sample comprises 88 PNe, several \HII\ regions and Wolf-Rayet shells, and an isolated filament of the Vela SNR.  For each object the SHS H$\alpha$ flux is measured following the recipe described in detail below, and is either compared with the SHASSA-derived \ha\ flux from FBP13, or with other independent \ha\ fluxes from the literature where available.  For the ejecta nebulae Hf\,39  and Wray\,15-751, the \ha\ fluxes, uncorrected for reddening, have been derived from the reddening-corrected fluxes and extinction values given in Stahl (1986, 1987), Hutsemekers \& van Drom (1991) and Hutsemekers (1994). 

\subsection{Measurement Process}\label{sec:measurement_pipeline}
Suitably sized H$\alpha$ and matching SR images from each SHS field that contain the selected calibrating objects were either downloaded as full 0.67\arcsec\,pix$^{-1}$ FITS images from the SHS website or extracted from the 16$\times$ blocked-down data depending on the object's angular size.  The SR image was then subtracted from the H$\alpha$ equivalent in {\tt IRAF} using the scaling factor provided in Table~\ref{tab:main}.  This is straightforward thanks to the in-built World Coordinate System (WCS) for each field.  These continuum-subtracted \ha\ images were then used for flux measurements. 

\begin{figure*}
\begin{center}
\includegraphics[height=4.85cm]{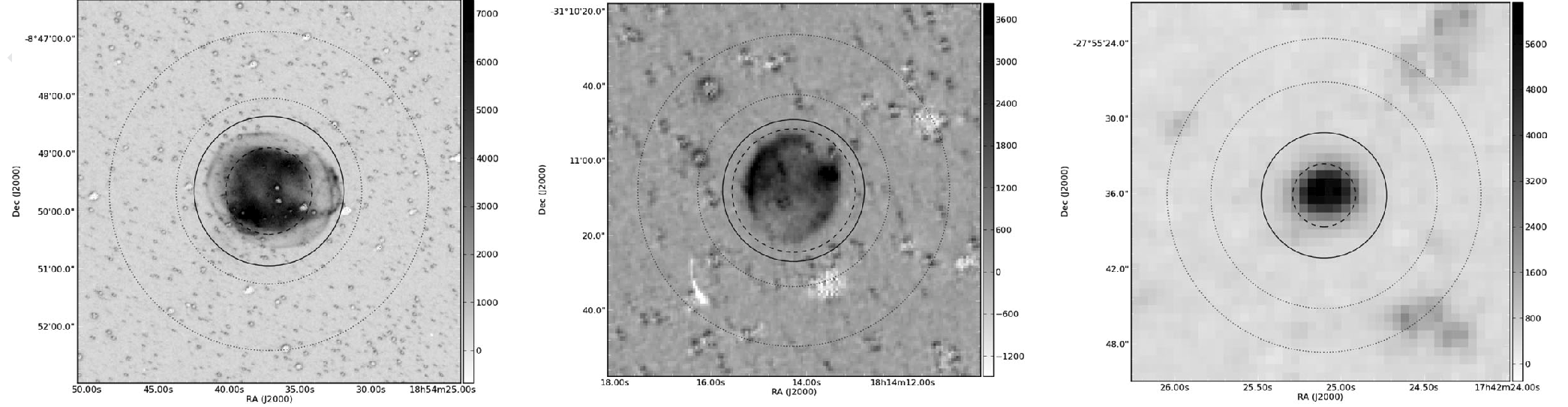}
\caption{SHS continuum-subtracted images showing the photometric apertures (solid circles) and background sky annuli (dotted circles) for three PNe:  ~IC~1295 (left), SB~4 (middle) and  the compact object JaSt~36 (right).  The nominal catalogued diameter of each PN is given by the dashed circle in each panel.  The images are 400\arcsec,  135\arcsec, and 50\arcsec\ on a side, respectively, and have NE at top left.   
} 
\label{fig:PN}
\end{center}
\end{figure*}

A photometric aperture was defined to carefully enclose the entire object in {\tt IRAF}, and sky background subtraction was achieved by using a suitably chosen sky annulus positioned to avoid, as far as possible, stellar residuals and other artefacts.  Unlike SHASSA, the superior resolution of the full-resolution SHS data often enables the fine structures and extended halos of most target sources to be seen.     
The background-corrected photometry counts were then used to measure the source \ha\ flux  using algorithms similar to those described in FBP13.   In Figure~\ref{fig:PN} we present two examples to illustrate the fitting procedure we employ to derive sky-subtracted integrated SHS counts for our calibrating sources. 

\begin{figure*}
\begin{center}
\includegraphics[height=4.85cm]{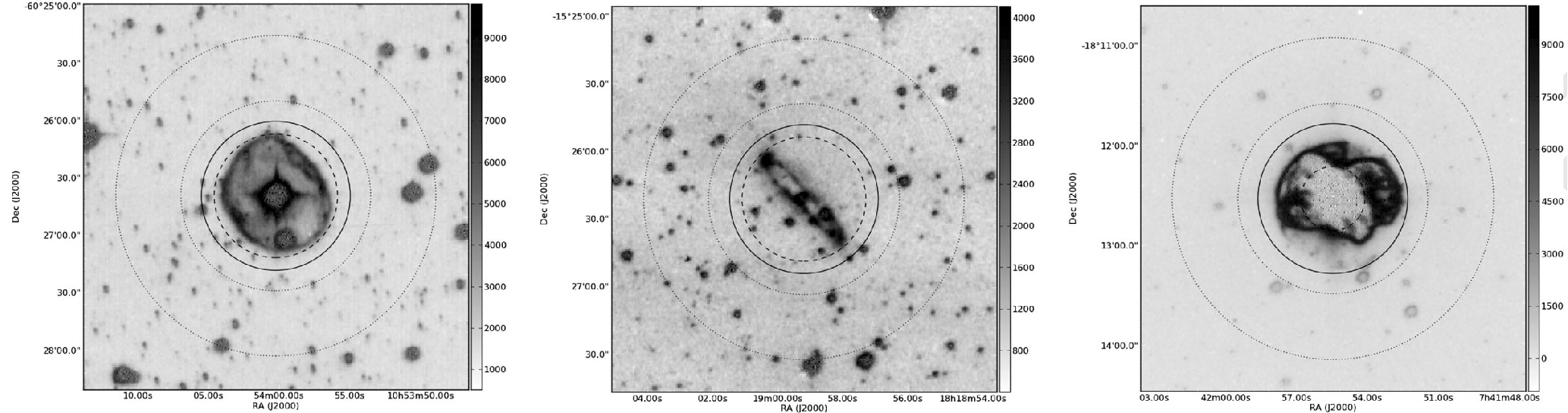}
\caption{SHS continuum-subtracted images illustrating the effect of a bright stellar residual from the massive WN11 star, Hen~3-519 (in its surrounding nebula, Hf~39, left), several residuals, including from the binary central star of the toroidal PN, PHR~J1818-1526 (middle), and a very bright, saturated PN with flux over-subtraction in its interior (NGC~2440, right).  The PN diameters, photometric apertures and background sky annuli are selected as before. The images are 170\arcsec, 320\arcsec, and 230\arcsec\ on a side, respectively, and have NE at top left. } 
\label{fig:PN2}
\end{center}
\end{figure*}

A problem often encountered is the presence of stellar residuals within the image area of extended objects.  In the Galactic plane, the stellar number density is high so residuals may also be present in the immediate region surrounding the chosen calibrator (Figure~\ref{fig:PN2}), making selection of suitable sky apertures more challenging.  Stellar residuals can also sometimes appear around bright, saturated stars or compact emission sources in the continuum subtracted products (Pierce 2005; Paper I).   Other challenges include residuals of the same angular size or larger than some compact PNe, which partly overlap the PN image. Although these saturated residuals were removed via the same method as above, there can be some contamination of the PN emission in a few cases.  
Residuals were present at some level in the images of most objects, but not all intensity contributions from these were corrected.  This was because the majority of the calibrating objects  studied (mainly PNe) were of moderate to large angular extent and consequently their images were populated by many small stellar residuals. Correcting for such large numbers of stellar residuals is difficult and time-consuming, and in many cases the effect of these residuals are accounted for quite effectively in the sky subtraction process.

Furthermore, recall the SR exposures also record H$\alpha$ emission, which, when subtracted from the matching H$\alpha$ exposures, can lead to an under estimation of the true H$\alpha$ flux.  Fortunately, In most cases this process appears to be self-compensating for the addition of H$\alpha$ emission by residuals, as our excellent calibration results testify (see below).  These results show that the integrated fluxes seem little affected for nebulae below an intensity limit of $\sim$7000 counts per pixel.  
If the contribution to the total H$\alpha$ emission of a particular source from stellar residuals seems significant, we attempt to correct for this so as to provide the most accurate possible continuum subtracted H$\alpha$ fluxes.  A correction was also  applied in the presence of a particularly bright PN central star, or if the calibrating object is located close to a bright, saturated star or stars being embedded in the object of interest.  In the case of Hf~39, the ionizing star is prominent relative to the nebula (Figure~\ref{fig:PN2}).  We simply measured the flux of the star with a small aperture and subtracted this from the total flux. This technique can also be applied to estimate fluxes for faint, extended PNe containing emission-line central stars (e.g. DePew et al. 2011).

\subsubsection{Deconvolution of \NII\ Emission}\label{sec:deconvolve}
Another factor that must be considered in determining accurate H$\alpha$ fluxes is  the close proximity of the  \NII\ $\lambda$6548\AA\ and $\lambda$6584\AA\ emission lines either side of H$\alpha$, which are included in the SHS  filter bandpass (Paper~I).  This is the case for most Galactic sources of modest radial velocity.  The contribution from the \NII\ lines to the total signal through the filter can vary significantly among emission sources. For example, for Type~I PNe (Peimbert 1978), the \NII/H$\alpha$ ratio is usually more than three and can sometimes exceed 10 (Corradi et al. 1997; Perinotto \& Corradi 1998; Kerber et al. 1998; Tajitsu et al. 1999; Frew, Parker \& Russeil 2006).  This must be taken into account when comparing SHS \ha\ fluxes with those determined independently in the literature.  This can be easily done when slit spectroscopy is available as this provides the observed \NII/H$\alpha$ ratio.  We have chosen calibrators that have spectra available, so the  \NII/H$\alpha$ ratios are known.  These are taken from the literature sources listed in FBP13, or elsewhere.  The observed ratios, usually obtained from long-slit spectroscopy, are deemed representative of the whole object.  This is a safe assumption in most cases, though significant internal variation across some PNe and SNRs has been observed.

The calculation of the SHS filter transmission properties is made more difficult by the blue shifting of the band pass with respect to incident angle in the fast f/2.48 converging beam of the UKST (see Parker \& Bland-Hawthorn 1998). The effect was modelled by breaking down the contribution from the converging light cone into a series of concentric rings (see Paper~I). The responses from each ring of the beam are weighted according to the areas of the ring and summed to generate a smeared filter response. This allows the calculation of the contributions from the \NII\ lines to the total flux recorded by the survey.   The reader is referred to Figure~3 of Paper I for an illustration of the modelled H$\alpha$ filter transmission profile.  Based on this profile, the SHS filter transmits H$\alpha$ at 80\% and \NII\ $\lambda$6548 and $\lambda$6584\AA\ at 82\% and 50\% respectively.  Given that the intrinsic ratio between the $\lambda$6548\AA\ and $\lambda$6584\AA\ \NII\ line strengths is quantum mechanically fixed at $\sim$3.0, the SHS filter transmits approximately 58\% of \NII, compared to 80\% transmission of the H$\alpha$ flux.  This gives a  \NII/H$\alpha$ filter transmittance ratio of 0.725, which is the constant in equation~\ref{eq:shs_deconvolve}, below.  

While the filter transmission falls towards the off-axis edge of each survey field (see Figure~\ref{fig:HAL0970}), this is not considered significant in this study due to the purposely large overlap between adjacent SHS fields on a 4\arcdeg\ grid (Paper~I).  This enables the data of interest to be preferentially selected from the non-vignetted 3\arcdeg\ $\times$ 3\arcdeg\ central region of an available field, where the transmission is better than 95\% relative to the peak.  The filter transmission constant will vary at large off-axis angular distances but this is only relevant for the few fields at the eastern edges of the SHS region.

\subsection{SHS \ha\ Fluxes}\label{sec:SHS_fluxes}
The equation used to derive the integrated SHS \ha+\NII\ (or `red') flux is  essentially the same as the expression given in FBP13.  The flux in cgs units is given by:
\begin{equation}
\label{eq:flux}
F_{\rm red} = 5.66\times10^{-18}\times P_S^2\times \Big(\frac{\rm SUM}{\rm CF}\Big) \quad   {\rm erg}\,{\rm cm}^{-2}\,{\rm s}^{-1}
\end{equation}
where the first constant in the expression is the conversion factor from rayleighs to cgs units, $P_S$ is the `plate scale' of the digital SHS data (0.67\arcsec\ per pixel for the full resolution images, and 10.7\arcsec\ per pixel for the 16$\times$ blocked-down images).   The summed counts obtained from the aperture photometry procedure  (SUM) need to be divided by the relevant calibration factor, CF,  taken from table~\ref{tab:main}, in order to determine the integrated red flux in units of erg\,cm$^{-2}$\,s$^{-1}$.   The SHS field number can be obtained from the header of any downloaded FITS image. 

The integrated H$\alpha$ flux is then obtained by correcting for the contribution from the contaminant \NII\ lines (Frew 2008; FBP13) and is given  by the following expression:
\begin{equation}
\label{eq:shs_deconvolve}
F({\rm H}\alpha)_{\rm SHS} = \frac{F_{\rm red}}{K_{\rm tr}\,R_{\rm \NII} + 1}
\end{equation}

where $R_{[\rm N\,II]}$ is the observed spectroscopically determined \NII/\ha\ ratio for the emission object in question (mainly PNe)  and the constant, $K_{\rm tr}$ = 0.725, takes into account the throughput of the SHS filter.   Following FBP13, the \NII\ flux refers to the sum of the $\lambda$6548 and $\lambda$6584 lines.  If only the brighter $\lambda$6584 line is measurable in a spectrum, $F$\NII\ is estimated as 1.33 $\times$ $F$($\lambda$6584) as the 6584 to 6548\AA  \NII\ line ratio is 3:1.   

The uncertainty budget for the SHS \ha\ flux is determined following the procedure presented in FBP13.   Additionally, the uncertainty in the CF for the particular SHS field, as measured from the fit,  was added in quadrature to the uncertainty in aperture intensity counts.  
We emphasise that the \NII\ transmission differs between the SHS and the SHASSA survey filters (70\AA\ and 32\AA\ FWHM respectively), so the SHS filter passes considerably more \NII\ emission.  The transmissible \NII/H$\alpha$ ratios of the two survey filters are 0.725 and 0.375 respectively.  In the original calibration of the SHS from SHASSA data (Paper I), both \HII\ regions and the diffuse ionized gas were assumed to have negligible \NII\ emission.  However, the median \NII/H$\alpha$ ratio for a large number of Galactic \HII\ regions is $\sim$0.4 (see the data graphically presented by Frew \& Parker 2010), which is in reasonable agreement with the average $\lambda$6584/\ha\ ratio of $\sim$0.3 seen in a sample of faint \HII\ regions (Madsen et al. 2006a).  These authors also noted that the $\lambda$6584/\ha\ ratio in the warm interstellar medium (WIM) generally increased below an \ha\ intensity of 10\,R, approaching unity at the faintest measured intensities.  Owing to these points, there is an offset of $\sim$10--15\,\% in the flux obtained from using the CF values presented in Table~\ref{tab:main}.  We therefore recommend that the SHS H$\alpha$ fluxes given by equation~\ref{eq:shs_deconvolve} be increased by this factor; i.e. the logarithmic H$\alpha$ fluxes should be brightened by 0.05\,dex.

The total uncertainty on the SHS \ha\ flux is given by the following expression, which includes a SHASSA uncertainty term as the SHS is bootstrapped to this survey:
\begin{equation}\label{eq:tot_uncertainty}
\sigma^2_{F {\rm (H\alpha)}} =  {\sigma}_{\rm phot}^{2} +  {\sigma}_{R}^{2} +  {\sigma}_{\rm cal}^2 +  \sigma_{\rm SHASSA}^{2} 
\end{equation}

where  $\sigma_{\rm phot}$ is the proportional uncertainty in the photometry measurement,  $\sigma_{R}$ is the uncertainty in the deconvolved \ha\ flux due to the uncertainty in $R_{\rm \NII}$,  $\sigma_{\rm cal}$ is the calibration uncertainty for the respective SHS field, and $\sigma_{\rm SHASSA}$ is the known SHASSA calibration uncertainty of 8\% (Gaustad et al. 2001). 

\subsubsection{Emission Measures}
The SHS pixel data can also be used in the study of the WIM and low-surface brightness \HII\ regions on spatial scales down to 2--3\arcsec, showing fine detail that is not seen on SHASSA.   For ionized hydrogen, the surface brightness, $S_{\rm H\alpha}$ (in R) and the emission measure (EM, in units of cm$^{-6}$\,pc) are related by the following expression (Haffner et al. 2003; Madsen, Reynolds \& Haffner 2006a; cf. Reynolds \& Ogden 1982):
\begin{equation}
\label{eq:em_measure1}
{\rm EM} = \int n_e^2\,dl = 2.75\,T^{0.9}\,S_{\rm H\alpha}\,e^{2.2 E(B-V)}\quad {\rm cm^{-6}\,pc}
\end{equation}

where $T$ is the electron temperature in units of 10$^{4}$\,K.  As before, the pixel counts can be converted into rayleighs using the CF for the specific SHS field, and then the contribution from the contaminant \NII\ lines needs to be deconvolved from the total measured intensity to determine an \ha\ surface brightness or emission measure (see \S\ref{sec:deconvolve}).  

The resulting data has wide application to the study of the structure and ionization of \HII\ regions and the WIM (Sembach et al. 2000; Reynolds 2004; Cox 2005), and can be readily compared with current and planned imaging surveys at other wavelengths across the electromagnetic spectrum, particularly at radio (e.g. Green et al. 1999; Murphy et al. 2007; Norris et al. 2011; Hoare et al. 2012) and infrared wavelengths (Churchwell et al. 2009).

\section{Comparison of SHS H$\alpha$ Fluxes against Independent Estimates}

Table~\ref{calib} gives our measured SHS H$\alpha$ fluxes for selected PNe and other calibrating emission-line sources.  The fluxes are based on our newly determined individual field calibrations.  We also list the independent integrated \ha\ fluxes for these objects.  The SHASSA \ha\ fluxes are primarily taken from FBP13, or measured here in the same manner, while \ha\ fluxes for the other nebulae are drawn from Stahl (1987), Hutsemekers (1994), Nota et al. (1995), Duerbeck \& Benetti (1996),  Weinberger, Kerber  \& Gr\"obner (1997), Pollacco (1999), Beaulieu, Dopita \& Freeman (1999), Boumis et al. (2003, 2006), Jacoby \& Van de Steene (2004), Ruffle et al. (2004), Madsen et al. (2006b), Parker et al. (2011, 2012b), and Frew et al. (2014).  FBP13 showed that the uncertainties from Cappellaro et al. (2001) are substantial, so we have not used these in our comparison.  


\begin{table*}
\begin{center}
\caption{A comparison of our SHS H$\alpha$ fluxes (in erg\,cm$^{-2}$\,s$^{-1}$) for 88 PNe and 13 other nebulae with independent integrated \ha\ fluxes taken from the literature, as explained in the text.  The table is published in its entirety as an online supplement, and a portion is shown here for guidance regarding its form and content.} 
\label{calib}
\begin{tabular}{lccccl}
\hline 
Name &~~log$F$(\ha)~~ &Reference &log$F$(\ha)  &SHS Field  & Remarks \\
  		&(other) &					&(SHS) &			&			\\
\hline \noalign{\smallskip}
Abell 18	&	~~$	-12.00	\pm	0.09	$~~	&	FBP13	&	$	-11.92	\pm	0.09$	&	~~HAL1197~~	&	$\ldots$	\\	
Abell 23	&	$	-11.86	\pm	0.09	$	&	FBP13	&	$	-11.82	\pm	0.08$	&	HAL0526	&	$\ldots$	\\	
Abell 26	&	$	-12.24	\pm	0.08	$	&	FBP13	&	$	-12.31	\pm	0.15$	&	HAL0601	&	$\ldots$	\\	
Abell 27	&	$	-12.01	\pm	0.05	$	&	FBP13	&	$	-12.01	\pm	0.15$	&	HAL0602	&	$\ldots$	\\	
Abell 44	&	$	-11.49	\pm	0.10	$	&	FBP13	&	$	-11.56	\pm	0.08$	&	HAL0970	&	$\ldots$	\\	
Abell 45	&	$	-11.26	\pm	0.10	$	&	FBP13	&	$	-11.35	\pm	0.15$	&	HAL1060	&	$\ldots$	\\	
Abell 48 	&	$	-11.58	\pm	0.06	$	&	FBP14	&	$	-11.55	\pm	0.10$	&	HAL1241	&	$\ldots$	\\	
Abell 58 	&	$	-12.28	\pm	0.06	$	&	FBP13	&	$	-12.32	\pm	0.11$	&	HAL1333	&	$\ldots$	\\	
BMP J1613-5406	&	$	-11.60	\pm	0.09	$	&	FBP13	&	$	-11.50	\pm	0.12$	&	HAL0289	&	$\ldots$	\\	
BMP J1808-1406	&	$	-11.68	\pm	0.09	$	&	FBP13	&	$	-11.78	\pm	0.14$	&	HAL0969	&	$\ldots$	\\	
\noalign{\smallskip}
CGMW 3-2111	&	$	-12.38	\pm	0.10	$	&	B06, FBP13&	$	-12.40	\pm	0.10$	&	HAL1058	&	$\ldots$	\\	
CVMP 1	&	$	-11.50	\pm	0.10	$	&	FBP13	&	$	-11.63	\pm	0.14	$&	HAL0232	&	$\ldots$	\\	
DPV 1	&	$	-12.5	\pm	0.3	$	&	DB96, P99	&	$	-12.50	\pm	0.11$	&HAL0968		&	Sakurai's object	\\	
FP J0711-2531 		&	$	-10.69	\pm	0.04	$	&	FBP12	&	$	-10.51	\pm	0.11$	&	HAL0756	&	$\ldots$	\\	
FP J1824-0319	&	$	-10.40	\pm	0.10	$	&	M06, FBP13	&	$	-10.44	\pm	0.10 $	&	HAL1240	&	$\ldots$	\\	
Fr 2-7	&	$	-10.82	\pm	0.10	$	&	FBP13	&	$	-10.86	\pm	0.12$	&	HAL0097	&	$\ldots$	\\	
G4.4+6.4	&	$	-11.25	\pm	0.06	$	&	FBP13	&	$	-11.31	\pm	0.10$	&	HAL0878	&	$\ldots$	\\	
GL1PN J1530-5557 	&	$	-13.86	\pm	0.10	$	&	PC12$^a$&	$	-14.17	\pm	0.12	$&	HAL0233	&	$\ldots$	\\	
GL1PN J1557-5430 	&	$	-14.79	\pm	0.15	$	&	PC12$^a$&	$	-14.70	\pm	0.15$	&	HAL0234	&	$\ldots$	\\	
GL1PN J1642-4453 	&	$	-14.94	\pm	0.15	$	&	PC12$^a$&	$	-14.88	\pm	0.15$	&	HAL0413	&	$\ldots$	\\	
$\vdots$       &	$\vdots$  &	$\vdots$  		&	$\vdots$  		&	$\vdots$  & $\vdots$\\
\hline 
\end{tabular}
\flushleft
\noindent{\small Reference codes:
 ~ASTR91 -- Acker et al. (1991);   B03, B06 -- Boumis et al. (2003, 2006); BDF99 -- Beaulieu et al. (1999); DB96 -- Duerbeck \& Benetti (1996);  FBP13 --  Frew et al. (2013); FBP14 --  Frew et al. (2014);  H94 -- Hutsemekers (1994); HDM98 -- Hua, Dopita \& Martinis (1998); JaSt04 -- Jacoby \& Van de Steene (2004); KH93 -- Kistiakowsky \& Helfand (1993);  M06 -- Madsen et al. (2006b); P99 -- Pollacco (1999);  PC12 -- Parker et al. (2012b);  PFM11 -- Parker et al. (2011); RZ04 -- Ruffle et al. (2004);  S87 -- Stahl (1987);  SK89 -- Shaw \& Kaler (1989); WKG97 -- Weinberger et al. (1997); XP94 -- Xilouris et al. (1994);  XPPT -- Xilouris et al. (1996).~~ Note:  ~$^a$ Fluxes revised using updated calibration factors from this work.
}
\end{center}
\end{table*}

\begin{figure}
\begin{center}
\includegraphics[width=8.3cm]{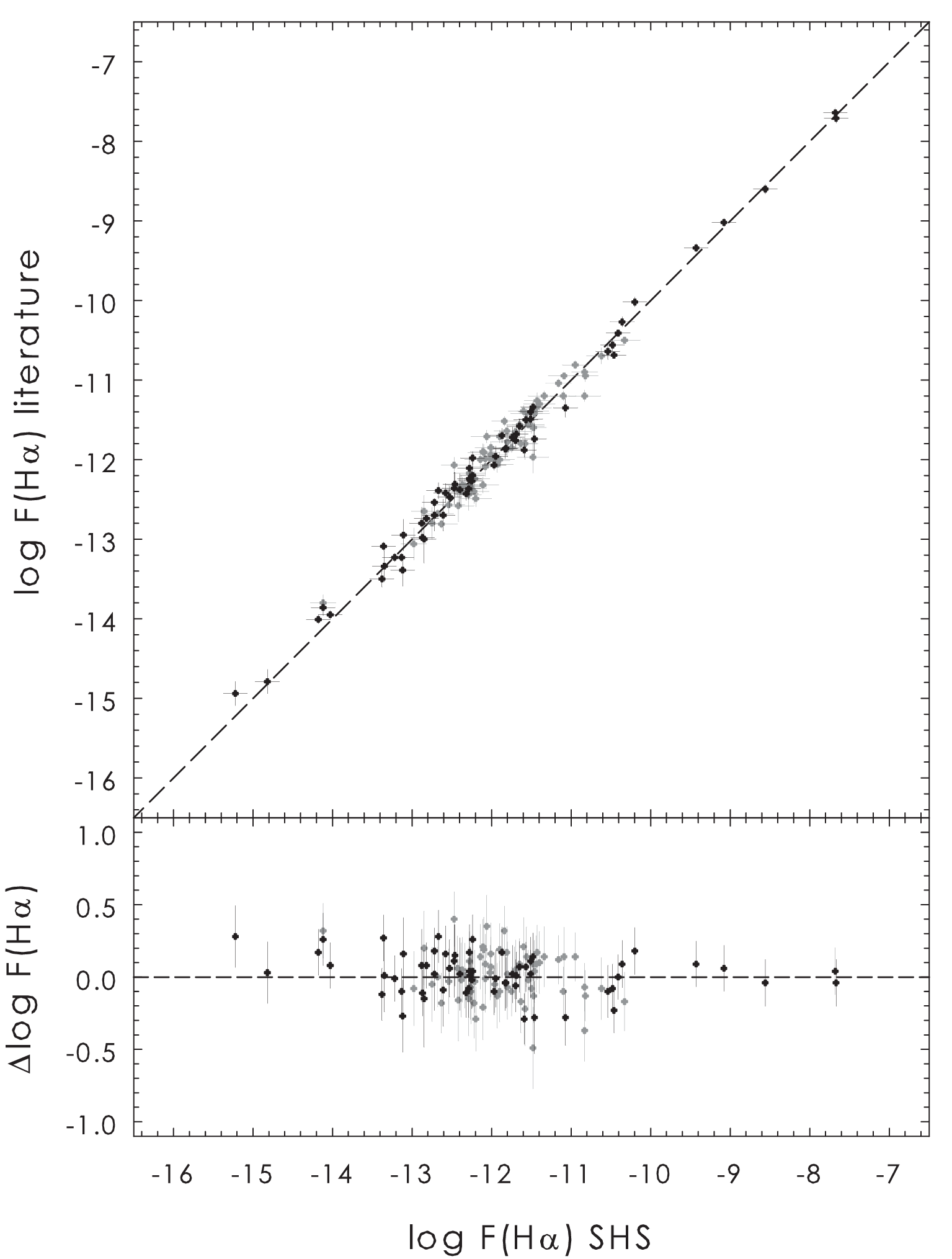}
\caption{SHS versus literature H$\alpha$ fluxes for the whole sample of unsaturated calibrating sources.  Fluxes derived from first- and second-ranked fields are plotted with filled and open circles, respectively.  The dashed line is the 1:1 relation.  Note that the agreement is excellent over a range in flux of more than 10 million. The lower panel plots the delta flux, in the sense of SHS minus literature fluxes.}
\label{fig:fluxcal}
\end{center}
\end{figure}

Figure~\ref{fig:fluxcal} presents our SHS \ha\ fluxes plotted against independent measurements from the literature for all 101 objects in the calibrating sample.  A tight linear relation is obtained over a range of 7\,dex (10 million) in integrated \ha\ flux.  The lower panel plots the logarithmic flux difference (in the sense of SHS minus literature) against the SHS flux.  Unsurprisingly the first-ranked fields permit the measurement of the most accurate fluxes, and the accuracy is slightly less for second ranked fields.  Our fluxes show a dispersion with the literature data of 0.12 dex from first-ranked fields, and 0.15 dex for second-ranked fields.  Considering the uncertainties on the literature fluxes, this translates to nominal uncertainties of 0.10 dex and 0.14 dex, respectively.  Thus, photographic data, if carefully treated, is a useful source for emission-line intensities and fluxes, accurate to $\sim$25--35 per cent, as seen for the flux data of Abell (1966), as reduced by FBP13.

\begin{figure}
\begin{center}
\includegraphics[width=8.3cm]{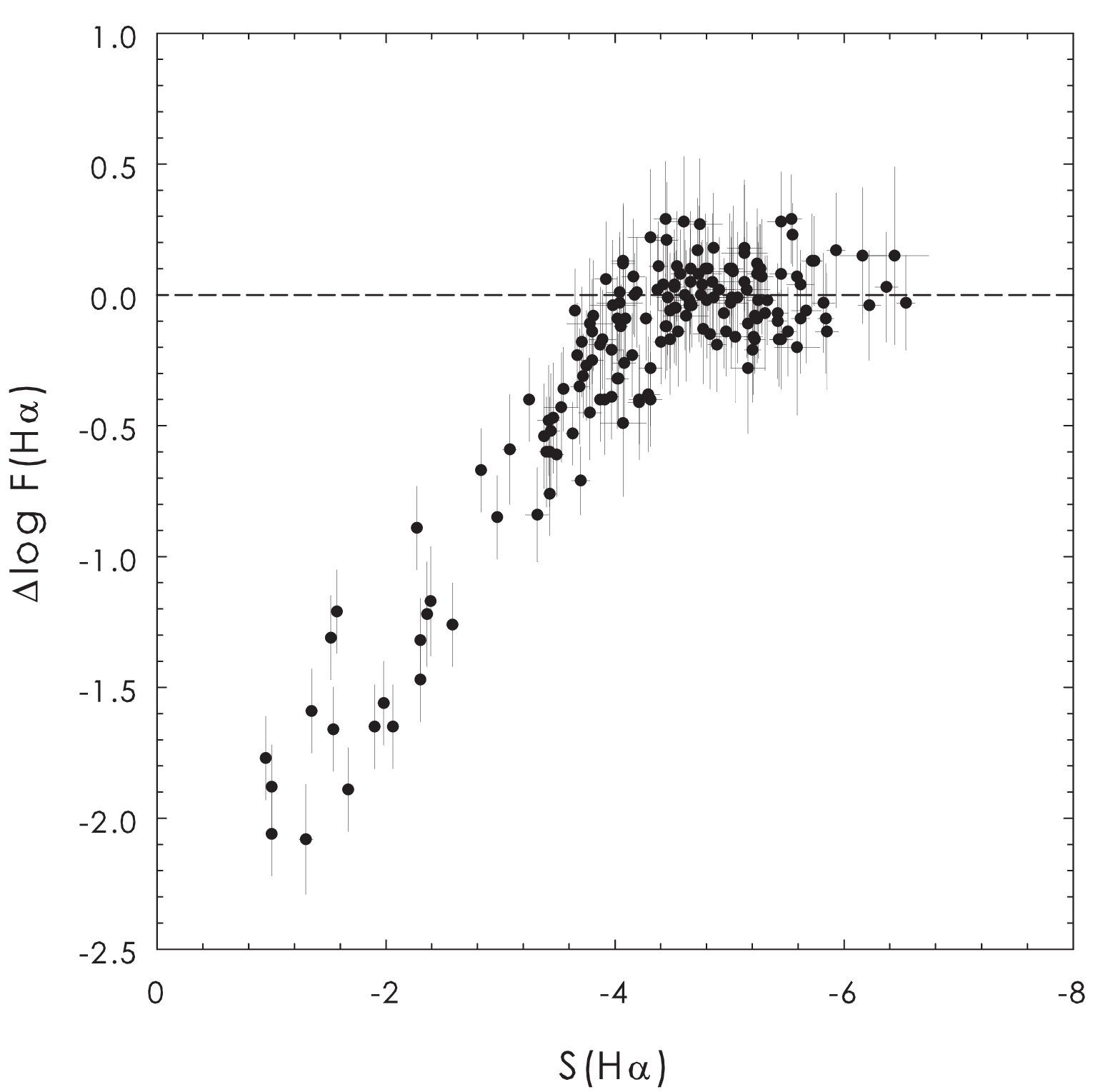}
\caption{The difference between SHS and literature H$\alpha$ fluxes plotted as a function of \ha\ surface brightness (in \ergcms\,sr$^{-1}$) derived from the literature flux and diameter for each object.  The twin effects of photographic saturation and flux over-subtraction occur for the brighter nebulae with a surface brightness, log$S$(\ha) $> -4.0$ \ergcms.}
\label{fig:saturation}
\end{center}
\end{figure}

In Figure~\ref{fig:saturation}, we show the offset between the SHS and literature H$\alpha$ flux plotted as a function of \ha\ surface brightness (in \ergcms\ sr$^{-1}$), in turn derived from the literature flux and the angular size of the object.  The sample is comprised of the objects from Table~\ref{calib}, plus another 45 saturated PNe, including very bright PNe like NGC\,3918 and NGC\,2440 (Figure~\ref{fig:PN2}).  The fluxes for these PNe are taken from FBP13 (and references therein) and the angular diameters are from Tylenda et al. (2003), Frew (2008), or the references given in the footnotes to Table~\ref{calib}.  The twin effects of photographic saturation and flux over-subtraction occur for the brighter nebulae with a surface brightness, log\,$S$(\ha) $\geq -4.0$, where an inflection point in the plot is clearly seen.

In a follow-up paper (Boji{\v c}i{\'c} et al., in preparation), we will provide new H$\alpha$ fluxes for all measurable PNe found in the SHS footprint, that were to faint to be measured from SHASSA.  These fluxes will complement the recent flux catalogue of FBP13.  In particular, the new SHS H$\alpha$ fluxes for faint Galactic bulge PNe  will be valuable, as they can be compared with new radio continuum fluxes (e.g. Boji\v{c}i\'c et al. 2011a,b) to estimate extinction values, or be compared with integrated fluxes in other wavebands; e.g. the \OIII\ fluxes of Kovacevic et al. (2011), in order to investigate the faint end of the bulge PN luminosity function in different emission lines (e.g. Ciardullo 2010).

\section{Point Source Photometry}
For completeness, we give a brief account of the SHS as a provider of photometric data for point-source emitters. Each \ha\ and matching SR survey field has been automatically processed through the standard SuperCOSMOS photometric and image processing pipelines  (Hambly et al. 1998, 2001). These provide so-called IAM (image-analysis mode) data for detected point-source and compact non-stellar sources in each survey passband above a locally thresholded background. An image-deblending algorithm, based on areal profile fitting at image levels equally spaced in log\,(intensity), is also applied (Beard, McGillivray \& Thanisch 1990; Hambly, Irwin \& MacGillivray 2001).  This is useful in the crowded star fields of the Galactic plane. Consequently, large catalogues of detected sources, including accurate positions, photometry and other image parameters are provided for each field from the SHS website\footnote{\url{http://www-wfau.roe.ac.uk/sss/halpha/}}.  A significant advantage of the SHS data is the ability to detect point sources that have been photometrically calibrated to CCD standards (Boyle, Shanks \& Croom 1995; Croom et al. 1999).  The narrowband \ha\ images are also calibrated to an `$R$-equivalent' scale. 

The 3-hour \ha\ and 15-min broad-band SR exposure times are chosen so that both reach similar depths of $R_S$ $\simeq$ 20.5 for point sources (Paper~I).  Where an object is detected in one band but not in the other, a default value of 99.999 is given in the online IAM catalogue data for the magnitude in the missing bandpass. Positional and magnitude-dependent errors are corrected for in the data that are available through the SHS website. Photometric consistency is achieved by using the field overlap regions to establish zero-points across the survey. These corrected magnitudes provide a means, for example, of selecting point-source \ha\ emitters by looking for significant brightening on the \ha\ source compare to its SR counterpart, especially when combined with the SSS $I$-band photometry which effectively eliminates late-type stellar contaminants (Paper~I).  The IAM data was successfully used by Pierce (2005) to discover a range of emission-line stars, by Hopewell et al. (2005) who found several WO and WC stars, and in the discovery of numerous compact PNe by Miszalski et al. (2008).

Modest variations in measured IAM stellar parameters from SHS data occur as a function of field position and the resulting variable PSF due to the effects of field rotation on the sky. This is apparent when comparing point sources taken on the  long 3\,hour \ha\ exposures with the short 15\,min SR exposures, and also from field vignetting, especially at large radii from the field centres. To better account for such effects requires that photometric comparisons are performed over limited 1\arcdeg\ areas such that  the true \ha\ emitters show a bona-fide enhanced \ha\ magnitude compared with the continuum SR magnitude.   At the bright end of the magnitude distribution, severe saturation effects also come in to play, limiting stellar photometry to $R$ of about $\sim$12 in both the \ha\ and SR passbands.  Further details are given in Paper~I.
\section{Conclusions and Recommendations}
We have established calibration factors (Table~\ref{tab:main}) for 170 of the 233  SHS fields (73\,\%), so these can be used directly to derive accurate integrated H$\alpha$ fluxes for a wide range of objects including PNe, \HII\ regions, Wolf-Rayet shells and supernova remnants (refer to Figure~\ref{fig:fluxcal}).  Based on the significant number of calibrating objects we have used, we can demonstrate that our derived fluxes agree to $\pm$0.1\,dex ($\approx$25 per cent) with independent fluxes from the literature over a remarkable range of 7\,dex in integrated flux.  There is a small systematic bias where SHS fluxes seem to be approximately 10--15\% fainter than the equivalent SHASSA fluxes.   The origin of this bias is attributed to the differing filter transmission of the \NII\ lines between SHASSA and the SHS, and the fact that the original analysis assumed that the diffuse emission from the ISM was solely \ha\ emission.  Better agreement  can be obtained by scaling up the SHS H$\alpha$ fluxes by 0.05\,dex. 

Determining fluxes for extended objects within the first-ranked SHS fields can be performed with confidence.   For most of the second-ranked fields,  the flux uncertainties are somewhat larger, though still usable.  There are currently 63 relatively featureless fields in Table~\ref{tab:main}) for which we could not determine the CF.   These fields either have high chemical fog levels or the fields have a poor dynamic range. This does not mean there is anything wrong per se with the exposures, just that the fields are mostly devoid of variable intensity \ha\ emission making it difficult to determine the CF.  Such fields may still contain significant numbers of small localised emission nebulae such as PNe.  In the meantime, we use the mean CF for all first-ranked fields of 12.0\,counts\,px$^{-1}$\,R$^{-1}$ as a default value for fields with indeterminate values.  The \ha\ surface brightnesses and integrated fluxes so derived should be accurate to better than $\pm$0.2\,dex (or $\approx$50 per cent).  Many of these fields still could be used with more confidence if local CF values were established, even for fields affected by high chemical fog.  

The adopted SF and CF values and rank status of each of the 233 SHS fields are given in Table~\ref{tab:main}, and is the main legacy material from this work.  The SHS fields can be used not just for exploratory science and discovering new astronomical objects, but now for deriving their integrated H$\alpha$ fluxes and surface brightnesses.  The SHS will remain useful for many years to come, and will be superseded by the new VPHAS+~\ha\ survey (Drew et al. 2014) only in the region along the SGP within $|b| \leq$ 5\arcdeg.  Even there, the SHS will be the source of first-epoch properties for emission-line stars and other exotica for future studies on variability.

In summary, the following notes and caveats should be considered when using the SHS data to provide \ha\ intensities and integrated flux measurements:
\begin{enumerate}[1.]
\item The pixel data has been treated by a filter flat-field as described in Paper I.  This is a one-off correction that is assumed valid for all survey fields.
\item We recommend that objects located more than 1.8\arcdeg\ from the SHS field centre not be measured for flux if accuracy is a requirement.  Excepting for the fields on the east, south and northern edges of the SHS zone, the generous overlap between field centres means that a neighbouring field should be available for flux measurement.  Note that when full 0.67\arcsec\,pix$^{-1}$ resolution areas are obtained from the SHS webpage using the ``Get Image'' tool, the best data are automatically selected from the unvignetted central region of an available field.
\item The effect of geocoronal \ha\ emission should be low.  Paper I found from a random sample of six SHS fields that the Earth shadow height for the whole 3-hour observation and across the 5\arcdeg\ field of view was greater than 6000\,km, so low-level geocoronal emission ($<$2\,R) is not problematic, as most fields covered by the SHS will contain significantly stronger diffuse emission. 
\item Telluric OH emission contaminates the images at low flux levels, but this is also altitude dependent and can be safely ignored.
\item The image data are non-linear at high intensities. Photographic saturation sets in at around 500 -- 600\,R, or approximately 6000 -- 7000 counts per pixel (for a typical CF value).  Objects showing values above this, or showing `negative' residuals after continuum subtraction should be treated with caution, and the total fluxes can be considered to be firm lower limits. Consider point-source algorithms if the nebula is bright \textit{and} compact. 
\item Brighter nebulae that are too extended for the application of point-source algorithms should be conspicuous enough to be detected in SHASSA.  We recommend the use of this survey in such cases. 
\item Emission from the red \NII\ lines often contributes significantly to the \ha+\NII\ flux, especially for PNe and SNRs (Frew \& Parker 2010; Sabin et al. 2013).  This contribution needs to be carefully de-convolved from the total flux using the observed spectroscopic ratio for individual objects,  and following the procedure described in \S\ref{sec:SHS_fluxes}.
\item Owing to the different passbands of the SHASSA and SHS filters, the SHS H$\alpha$ fluxes determined from Equation\,\ref{eq:shs_deconvolve} show a small systematic bias, and should be increased by an additional factor; i.e. the logarithmic H$\alpha$ fluxes should be brightened by 0.05\,dex. 
\item Integrated fluxes accurate to $\pm$\,0.10\,--\,0.14\,dex ($\sim$25--35 per cent), can be generally obtained across the 170 survey fields for which have a determined a CF, with the second-ranked fields have the larger systematic errors. 
\item A default value of 12.0 $\pm$ 3.0 counts\,px$^{-1}$\,R$^{-1}$ can be used as the CF for the 63 SHS fields with no formal determination of this parameter (see Figure~\ref{fig:coverage}).  The integrated \ha\ fluxes for any objects derived from these fields should be accurate to better than $\pm$\,0.2 dex (or $\approx$50 per cent). Alternatively, any unsaturated PNe or other small nebulae with an independent flux can be used to calibrate the whole field.  In such cases, the CF value for the field can be treated as a free parameter and varied until the measured integrated SHS flux matches the independent value.
\item The nature of photography and the scanning process produces a range of artefacts and spurious images on the SHS.  The user should be aware of these, and can find more details in Paper~I.
\end{enumerate}

\section*{Acknowledgments} 
This research used data from the AAO/UKST \ha\ Survey, produced with the support of the Anglo-Australian Telescope Board and the UK Particle Physics and Astronomy Research Council (now the STFC), and data from the Southern \ha\ Sky Survey Atlas (SHASSA), which was produced with support from the National Science Foundation.  This research made use of the SIMBAD database and the VizieR service, operated at the CDS, Strasbourg, and also made use of Montage, funded by the National Aeronautics and Space Administration's Earth Science Technology Office, Computation Technologies Project, under Cooperative Agreement Number NCC5-626 between NASA and the California Institute of Technology.  DJF is grateful to Macquarie University for the award of a Macquarie Research Fellowship, ISB thanks the ARC for his Super Science Fellowship (project ID FS100100019), and QAP acknowledges additional support from Macquarie University and the Australian Astronomical Observatory.

\newpage
\onecolumn
\section*{Supplementary Online Material}\label{appendix}

{\small
\begin{center}
\begin{longtable}{ccccl}
\caption{(TABLE\,1): Data summary for all 233 SHS survey fields.   Note there are 63 fields with no calibration factors determined (usually due to poor dynamic range or high chemical fog, or both).  A default value of 12.0 $\pm$ 3.0 counts\,pix$^{-1}$\,R$^{-1}$ can be used for these fields, as explained in the text.} 
\label{tab:main}\\
\hline \noalign{\smallskip}
Field &
Cont. factor &
Cal. factor  &
Quality  &
Comment  \\
\noalign{\smallskip}
 &
&
(count\,pix$^{-1}$R$^{-1}$)   & 
&
 \\
\hline \noalign{\smallskip}
\endfirsthead
\multicolumn{5}{c}{{\tablename} \thetable{} -- Continued} \\[0.5ex]
\hline \noalign{\smallskip}
Field &
Cont. factor &
Cal. factor  &
Quality &
Comment  \\
\noalign{\smallskip}
 &
&
(count\,pix$^{-1}$R$^{-1}$)    &
 &
\\ \hline 
\endhead
\hline \noalign{\smallskip}
\endfoot
\hline \noalign{\smallskip}
\endlastfoot
HAL0067 	&	...	&     (12.0)	&	3	&	small dynamic range	\\
HAL0068 	&	...	&	(12.0)	&	3	&	small dynamic range	\\
HAL0069 	&	...	&	(12.0)	&	3	&	small dynamic range	\\
HAL0070 	&	...	&	(12.0)	&	3	&	 photographic fog	\\
HAL0096 	&	...	&	13.3   	&	3	&	small dynamic range	\\
HAL0097	&	0.98	&	11.4	&	2	&	$\ldots$	\\
HAL0098 	&	0.83	&	13.9	&	2	&	 photographic fog	\\
HAL0099	&	0.42	&	(12.0)	&	3	&	 photographic fog	\\
HAL0100	&	1.10	&	12.7	&	1	&	$\ldots$	\\
HAL0101	&	1.12	&	11.7	&	2	&	small dynamic range	\\
HAL0102	&	...	&	(12.0)	&	3	&	small dynamic range	\\
HAL0103 	&	0.44	&	6.2	&	2	&	 photographic fog	\\
HAL0131 	&	0.62	&	(12.0)	&	3	&	 photographic fog	\\
HAL0132	&	0.91	&	10.0	&	2	&	$\ldots$	\\
HAL0133	&	1.29	&	6.3	&	2	&	$\ldots$	\\
HAL0134 	&	1.27	&	10.3	&	1	&	saturates at 500 R	\\
HAL0135 	&	0.92	&	6.6	&	1	&	$\ldots$	\\
HAL0136	&	0.87	&	12.0	&	2	&	$\ldots$	\\
HAL0137	&	1.24	&	18.9	&	2	&	large scatter	\\
HAL0138	&	0.91	&	8.7	&	2	&	large scatter	\\
HAL0139	&	0.90	&	13.2	&	2	&	large scatter	\\
HAL0140	&	0.96	&	10.0	&	2	&	large scatter	\\
HAL0141	&	1.16	&	9.5	&	2	&	large scatter	\\
HAL0171 	&	...	&	(12.0)	&	3	&	small dynamic range	\\
HAL0172 	&	...	&	(12.0)	&	3	&	 photographic fog	\\
HAL0173	&	...	&	16.3	&	2	&	small dynamic range	\\
HAL0174	&	1.06	&	12.7	&	2	&	$\ldots$	\\
HAL0175	&	0.65	&	11.2	&	1	&	saturates at 800 R	\\  
HAL0176	&	0.72	&	13.3	&	1	&	$\ldots$	\\
HAL0177	&	1.29	&	9.4	&	2	&	polynomial fit	\\
HAL0178	&	1.08	&	13.9	&	1	&	$\ldots$	\\
HAL0179	&	1.02	&	12.4	&	2	&	$\ldots$	\\
HAL0180 	&	1.05	&	13.2	&	2	&	$\ldots$	\\
HAL0181	&	1.42	&	10.8	&	2	&	poor fit	\\
HAL0182	&	1.12	&	7.0	&	2	&	$\ldots$	\\
HAL0183	&	1.00	&	13.1	&	2	&	$\ldots$	\\
HAL0184	&	1.04	&	7.2	&	1	&	$\ldots$	\\
HAL0185 	&	...	&	(12.0)	&	3	&	small dynamic range	\\
HAL0186 	&	...	&	(12.0)	&	3	&	small dynamic range	\\
HAL0219 	&	...	&(12.0)	&	3	&	small dynamic range	\\
HAL0220 	&	...	&	7.3	&	2	&	 photographic fog	\\
HAL0221	&	1.09	&	13.1	&	2	&	$\ldots$	\\
HAL0222	&	1.05	&	11.3	&	1	&	$\ldots$	\\
HAL0223	&	1.23	&	12.5	&	1	&	$\ldots$	\\
HAL0224	&	0.87	&	12.6	&	1	&	$\ldots$	\\
HAL0225	&	1.19	&	10.0	&	2	&	two trends	\\
HAL0226	&	1.12	&	(12.0)	&	3	&	small dynamic range	\\
HAL0227	&	1.28	&	(12.0)	&	3	&	small dynamic range	\\
HAL0228	&	1.09	&	6.6	&	1	&	$\ldots$	\\
HAL0229	&	1.42	&	13.9	&	2	&	$\ldots$	\\
HAL0230	&	...	&	10.0	&	2	&	small dynamic range	\\
HAL0231	&	0.96	&	(12.0)	&	3	&	 photographic fog	\\
HAL0232	&	1.23	&	12.6	&	2	&	$\ldots$	\\
HAL0233	&	1.14	&	9.9	&	2	&	$\ldots$	\\
HAL0234	&	0.99	&	12.8	&	1	&	$\ldots$	\\
HAL0235	&	0.64	&	15.0	&	2	&	 photographic fog	\\
HAL0236 	&	...	&	(12.0)	&	3	&	small dynamic range	\\
HAL0237 	&	1.05	&	11.4	&	2	&	 photographic fog	\\
HAL0272	&	0.96	&	11.1	&	2	&	$\ldots$	\\
HAL0273 	&	...	&	(12.0)	&	3	&	fit failed	\\
HAL0274 	&	1.29	&	12.8	&	1	&	$\ldots$	\\
HAL0275	&	1.07	&	14.3	&	2	&	$\ldots$	\\
HAL0276	&	0.47	&	(12.0)	&	3	&	photographic fog	\\
HAL0277	&	1.11	&	16.9	&	2	&	small dynamic range	\\
HAL0278 	&	1.13	&	18.0	&	2	&	poor fit	\\
HAL0279 	&	...	&	(12.0)	&	3	&	small dynamic range	\\
HAL0285 	&	0.41	&	13.9	&	2	&	photographic fog	\\
HAL0286 	&	...	&	(12.0)	&	3	&	photographic fog	\\
HAL0287 	&	0.44	&	(12.0)	&	3	&	photographic fog	\\
HAL0288 	&	1.32	&	(12.0)	&	3	&	small dynamic range	\\
HAL0289	&	1.15	&	18.3	&	2	&	 polynomial fit 	\\
HAL0290	&	0.68	&	12.0	&	2	&	poor fit	\\
HAL0291	&	0.65	&	17.9	&	2	&	photographic fog	\\
HAL0292	&	...	&	(12.0)	&	3	&$\ldots$		\\
HAL0329	&	0.87	&	11.6	&	2	&	two trends	\\
HAL0330 	&	0.63	&	13.1	&	1	&	$\ldots$	\\
HAL0331	&	1.06	&	7.3	&	2	&	$\ldots$	\\
HAL0332	&	1.08	&	9.3	&	2	&	$\ldots$	\\
HAL0333 	&	...	&	(12.0)	&	3	&	small dynamic range	\\
HAL0334 	&	0.74	&	(12.0)	&	3	&	 photographic fog	\\
HAL0335 	&	1.0	&	16.4	&	3	&	small dynamic range	\\
HAL0345 	&	1.25	&	(12.0)	&	3	&	 photographic fog	\\
HAL0346 	&	...	&	(12.0)	&	3	&	 photographic fog	\\
HAL0347 	&	1.09	&	10.7	&	2	&	 photographic fog	\\
HAL0348 	&	0.90	&	12.5	&	2	&	 polynomial fit 	\\
HAL0349 	&	1.06	&	18.5	&	1	&	 $\ldots$	\\    
HAL0350 	&	1.37	&	14.5	&	1	&	saturates at 500 R	\\
HAL0351 	&	0.78	&	11.5	&	1	&	$\ldots$	\\
HAL0352 	&	0.96	&	10.8	&	2	&	$\ldots$	\\
HAL0390 	&	1.03	&	9.0	&	1	&	$\ldots$	\\
HAL0391 	&	0.82	&	17.1	&	1	&	$\ldots$	\\
HAL0392 	&	0.89	&	15.5	&	1	&	$\ldots$	\\
HAL0393	&	1.07	&	9.9	&	2	&	poor fit	\\
HAL0394	&	1.04	&	10.2	&	2	&	$\ldots$	\\
HAL0395 	&	1.16	&	10.8	&	2	&	$\ldots$	\\
HAL0396 	&	1.07	&	8.5	&	1	&	$\ldots$	\\
HAL0410 	&	...	&	(12.0)	&	3	&	small dynamic range	\\
HAL0411	&	1.19	&	15.1	&	2	&	small dynamic range	\\
HAL0412 	&	1.05	&	(12.0)	&	3	&	small dynamic range	\\
HAL0413	&	0.83	&	11.3	&	2	&	large scatter	\\
HAL0414 	&	1.12	&	9.4	&	1	&	$\ldots$	\\
HAL0415	&	1.20	&	13.3	&	2	&	 polynomial fit 	\\
HAL0416 	&	1.27	&	14.7	&	2	&	$\ldots$	\\
HAL0417 	&	...	&	16.2	&	2	&	poor fit	\\
HAL0456	&	0.78	&	12.1	&	2	&	polynomial fit 	\\
HAL0457 	&	1.06	&	8.8	&	1	&	$\ldots$	\\
HAL0458	&	0.98	&	9.0	&	2	&	polynomial fit 	\\
HAL0459 	&	1.00	&	8.7	&	1	&	 $\ldots$	\\
HAL0460 	&	1.14	&	9.5	&	1	&	small dynamic range	\\
HAL0461	&	1.22	&	9.4	&	2	&	$\ldots$	\\
HAL0462 	&	1.02	&	12.8	&	1	&	 fog on SR	\\
HAL0480 	&	1.05	&	(12.0)	&	3	&	small dynamic range	\\
HAL0481 	&	1.27	&	14.2	&	1	&	$\ldots$	\\
HAL0482 	&	1.18	&	17.5	&	1	&	$\ldots$	\\
HAL0483 	&	0.91	&	13.1	&	1	&	$\ldots$	\\
HAL0484	&	0.56	&	11.7	&	2	&	$\ldots$	\\
HAL0485	&	...	&	14.0	&	2	&	small dynamic range	\\
HAL0526 	&	0.86	&	12.4	&	1	&	$\ldots$	\\
HAL0527	&	1.10	&	10.3	&	2	&	polynomial fit	\\
HAL0528	&	0.83	&	15.2	&	1	&	$\ldots$	\\
HAL0529	&	1.11	&	12.2	&	2	&	$\ldots$	\\
HAL0530	&	1.01	&	11.9	&	2	&	large scatter	\\
HAL0531	&	0.99	&	15.0	&	2	&	poor fit	\\
HAL0552	&	1.10	&	13.7	&	2	&	small dynamic range	\\
HAL0553	&	1.03	&	10.3	&	2	&	poor fit	\\
HAL0554 	&	1.17	&	16.2	&	1	&	$\ldots$	\\
HAL0555 	&	1.17	&	14.6	&	1	&	saturates at 600 R	\\
HAL0556 	&	1.01	&	14.0	&	2	&	$\ldots$	\\
HAL0557 	&	0.44	&	17.9	&	2	&	photographic fog	\\
HAL0598	&	...	&	10.2	&	2	&	large scatter	\\
HAL0599 	&	1.10	&	7.5	&	1	&	$\ldots$	\\
HAL0600	&	1.18	&	8.5	&	2	&	$\ldots$	\\
HAL0601	&	1.10	&	8.2	&	2	&	$\ldots$	\\
HAL0602	&	1.04	&	9.2	&	2	&	poor fit	\\
HAL0603	&	1.13	&	12.3	&	2	&	poor fit	\\
HAL0627	&	...	&	6.0	&	2	&	photographic fog	\\
HAL0628	&	1.30	&	15.3	&	2	&	two trends	\\
HAL0629 	&	1.04	&	12.0	&	3	&	small dynamic range	\\
HAL0630	&	0.99	&	14.3	&	2	&	two trends	\\
HAL0631 	&	1.16	&	12.8	&	1	&	$\ldots$	\\
HAL0632 	&	1.20	&	16.1	&	2	&	poor fit	\\
HAL0633 	&	...	&	(12.0)	&	3	$\ldots$&		\\
HAL0675	&	0.95	&	11.7	&	2	&	poor fit	\\
HAL0676	&	0.97	&	6.9	&	2	&	$\ldots$	\\
HAL0677	&	1.14	&	9.3	&	1	&	$\ldots$	\\
HAL0678	&	1.10	&	11.0	&	2	&	poor fit	\\
HAL0679 	&	1.21	&	(12.0)	&	3	&	small dynamic range	\\
HAL0680	&	1.08	&	12.9	&	2	&	poor fit	\\
HAL0707	&	1.19	&	15.4	&	2	&	polynomial fit	\\
HAL0708	&	1.18	&(12.0)	&	3	&	small dynamic range	\\
HAL0709	&	0.92	&	14.1	&	2	&	poor fit	\\
HAL0710 	&	0.97	&	13.4	&	1	&	$\ldots$	\\
HAL0711 	&	0.64	&	(12.0)&	3	&	 photographic fog	\\
HAL0712 	&	0.40	&	20.6	&	2	&	poor fit	\\
HAL0713 	&	1.10	&	14.2	&	2	&	 photographic fog	\\
HAL0756	&	1.23	&	13.0	&	1	&	$\ldots$	\\
HAL0757 	&	0.96	&	11.2	&	1	&	$\ldots$	\\
HAL0758	&	1.19	&	7.8	&	2	&	poor fit	\\
HAL0759 	&	1.15	&	6.6	&	2	&	poor fit	\\
HAL0760 	&	...	&	(12.0)	&	3	&	 photographic fog	\\
HAL0761	&	0.98	&	8.9	&	2	&	poor fit	\\
HAL0792	&	0.99	&	(12.0)	&	3	&	 photographic fog	\\
HAL0793	&	1.21	&	17.8	&	2	&	poor fit	\\
HAL0794 	&	0.73	&	11.1	&	1	&	$\ldots$	\\
HAL0795 	&	1.01	&	12.3	&	1	&	saturates at 700 R	\\
HAL0796	&	1.21	&	15.0	&	2	&	poor fit	\\
HAL0797 	&	1.11	&	(12.0)	&	3	&	 photographic fog	\\
HAL0841	&	1.13	&	9.4	&	2	&	small dynamic range	\\
HAL0842 	&	1.06	&	7.8	&	2	&	$\ldots$	\\
HAL0843	&	0.99	&	9.7	&	2	&	$\ldots$	\\
HAL0844 	&	...	&	9.7	&	2	&	 photographic fog	\\
HAL0845	&	1.15	&	14.9	&	2	&	poor fit	\\
HAL0878	&	0.99	&	12.3	&	2	&	poor fit	\\
HAL0879 	&	1.22	&	13.3	&	2	&	poor fit	\\
HAL0880 	&	0.98	&	14.0	&	1	&	linear fit	\\
HAL0881	&	1.51	&	9.0	&	1	&	$\ldots$	\\
HAL0882	&	1.18	&	(12.0)	&	3	&	small dynamic range	\\
HAL0926 	&	...	&	(12.0)	&	3	&	 photographic fog	\\
HAL0927	&	1.03	&	9.9	&	2	&	poor fit	\\
HAL0928	&	1.22	&	13.3	&	2	&	small dynamic range	\\
HAL0929	&	1.18	&	14.3	&	2	&	polynomial fit	\\
HAL0930 	&	0.80	&	(12.0)	&	3	&	 photographic fog	\\
HAL0931 	&	0.94	&	(12.0)	&	3	&	 photographic fog	\\
HAL0967 	&	0.77	&	21.4	&	2	&	 photographic fog	\\
HAL0968	&	...	&	14.4	&	2	&	poor fit	\\
HAL0969	&	1.14	&	20.4	&	2	&	 large scatter	\\
HAL0970	&	0.87	&	14.0	&	1	&	saturates at 600 R	\\
HAL0971 	&	1.19	&	8.5	&	1	&	$\ldots$	\\
HAL0972 	&	0.40	&	(12.0)	&	3	&	 photographic fog	\\
HAL1016 	&	...	&	(12.0)	&	3	&	small dynamic range	\\
HAL1017	&	1.11	&	7.0	&	2	&	 polynomial fit 	\\
HAL1018	&	0.98	&	9.8	&	2	&	 polynomial fit 	\\
HAL1019	&	...	&	(12.0)	&	3	&	small dynamic range	\\
HAL1020 	&	...	&	(12.0)	&	3	&	small dynamic range	\\
HAL1021	&	...	&	(12.0)	&	3	&	small dynamic range	\\
HAL1058	&	1.21	&	14.1	&	2	&	 polynomial fit 	\\
HAL1059	&	1.02	&	14.4	&	2	&	 polynomial fit 	\\
HAL1060	&	0.97	&	16.1	&	1	&	$\ldots$	\\
HAL1061	&	1.20	&	16.0	&	2	&	$\ldots$ 	\\
HAL1062	&	1.00	&	(12.0)	&	3	&	 photographic fog	\\
HAL1063 	&	...	&	(12.0)	&	3	&	small dynamic range	\\
HAL1106	&	1.24	&	7.8	&	2	&	$\ldots$	\\
HAL1107	&	...	&	10.0	&	2	&	poor fit	\\
HAL1108	&	1.08	&	7.5	&	1	&	$\ldots$	\\
HAL1109	&	1.07	&	7.7	&	1	&	polynomial fit	\\
HAL1110	&	...	&	(12.0)	&	3	&	small dynamic range	\\
HAL1149 	&	1.11	&	(12.0)	&	3	&	small dynamic range	\\
HAL1150 	&	1.19	&	16.6	&	2	&	$\ldots$	\\
HAL1151 	&	1.15	&	12.6	&	2	&	$\ldots$	\\
HAL1152	&	1.31	&	11.1	&	2	&	small dynamic range	\\
HAL1153 	&	...	&	(12.0)	&	3	&	small dynamic range	\\
HAL1195 	&	...	&	(12.0)	&	3	&	small dynamic range	\\
HAL1196 	&	0.56	&	(12.0)	&	3	&	 photographic fog	\\
HAL1197	&	1.08	&	7.0	&	2	&	poor fit	\\
HAL1198 	&	0.55	&	(12.0)	&	3	&	 photographic fog	\\
HAL1199 	&	0.77	&	(12.0)	&	3	&	 photographic fog	\\
HAL1200 	&	...	&	(12.0)	&	3	&	 photographic fog	\\
HAL1239 	&	...	&	(12.0)	&	3	&	 photographic fog	\\
HAL1240 	&	1.24	&	14.2	&	2	&	poor fit	\\
HAL1241	&	1.09	&	14.0	&	2	&	poor fit	\\
HAL1242	&	1.08	&	16.0	&	2	&	poor fit	\\
HAL1243 	&	1.18	&	(12.0)	&	3	&	small dynamic range	\\
HAL1244 	&	...	&	(12.0)	&	3	&	small dynamic range	\\
HAL1285 	&	0.92	&	(12.0)	&	3	&	 photographic fog	\\
HAL1286 	&	0.97	&	12.8	&	1	&	$\ldots$	\\
HAL1287	&	1.23	&	14.0	&	2	&	$\ldots$	\\
HAL1288 	&	1.25	&	(12.0)	&	3	&	 photographic fog	\\
HAL1289 	&	...	&	(12.0)	&	3	&	 photographic fog	\\
HAL1330	&	1.28	&	12.0&	2	&	small dynamic range	\\
HAL1331	&	1.24	&	13.7	&	2	&	$\ldots$	\\
HAL1332	&	1.04	&	14.8	&	2	&	large scatter	\\
HAL1333 	&	1.59	&	14.5	&	2	&	small dynamic range	\\
HAL1334 	&	0.80	&	(12.0)	&	3	&	 photographic fog	\\
\end{longtable}
\end{center}
}

\newpage
\onecolumn

{\small
\begin{center}
\begin{longtable}{lccccl}
\caption{(TABLE 2):  A comparison of our SHS H$\alpha$ fluxes (in erg\,cm$^{-2}$\,s$^{-1}$) for over 100 PNe and other nebulae with independent integrated \ha\ fluxes taken from the literature, as explained in the text.} 
\label{calib}\\
\hline \noalign{\smallskip}
Name &
~~log$F$(\ha)~~ &
Reference &
~log$F$(\ha)~  &
Field  &
Remarks \\
\noalign{\smallskip}
  &
(other) &
&
(SHS) &
  &
  \\
\hline \noalign{\smallskip}
\endfirsthead
\multicolumn{6}{c}{{\tablename} \thetable{} -- Continued} \\[0.5ex]
\hline \noalign{\smallskip}
Name &
~~log$F$(\ha)~~ &
Reference &
~log$F$(\ha)~ &
Field  &
Remarks \\
\noalign{\smallskip}
  &
(other) &
&
(SHS) &
  &
  \\
\hline \endhead
\hline \noalign{\smallskip}
\endfoot
\hline \noalign{\smallskip}
\endlastfoot
Abell 18	&	~~$	-12.00	\pm	0.09	$~~	&	FBP13	&	$	-11.92	\pm	0.09$	&	~~HAL1197~~	&	$\ldots$	\\	
Abell 23	&	$	-11.86	\pm	0.09	$	&	FBP13	&	$	-11.82	\pm	0.08$	&	HAL0526	&	$\ldots$	\\	
Abell 26	&	$	-12.24	\pm	0.08	$	&	FBP13	&	$	-12.31	\pm	0.15$	&	HAL0601	&	$\ldots$	\\	
Abell 27	&	$	-12.01	\pm	0.05	$	&	FBP13	&	$	-12.01	\pm	0.15$	&	HAL0602	&	$\ldots$	\\	
Abell 44	&	$	-11.49	\pm	0.10	$	&	FBP13	&	$	-11.56	\pm	0.08$	&	HAL0970	&	$\ldots$	\\	
Abell 45	&	$	-11.26	\pm	0.10	$	&	FBP13	&	$	-11.35	\pm	0.15$	&	HAL1060	&	$\ldots$	\\	
Abell 48 	&	$	-11.58	\pm	0.06	$	&	FBP14	&	$	-11.55	\pm	0.10$	&	HAL1241	&	$\ldots$	\\	
Abell 58 	&	$	-12.28	\pm	0.06	$	&	FBP13	&	$	-12.32	\pm	0.11$	&	HAL1333	&	$\ldots$	\\	
BMP J1613-5406	&	$	-11.60	\pm	0.09	$	&	FBP13	&	$	-11.50	\pm	0.12$	&	HAL0289	&	$\ldots$	\\	
BMP J1808-1406	&	$	-11.68	\pm	0.09	$	&	FBP13	&	$	-11.78	\pm	0.14$	&	HAL0969	&	$\ldots$	\\	
\noalign{\smallskip}
CGMW 3-2111	&	$	-12.38	\pm	0.10	$	&	B06, FBP13&	$	-12.40	\pm	0.10$	&	HAL1058	&	$\ldots$	\\	
CVMP 1	&	$	-11.50	\pm	0.10	$	&	FBP13	&	$	-11.63	\pm	0.14	$&	HAL0232	&	$\ldots$	\\	
DPV 1	&	$	-12.5	\pm	0.3	$	&	DB96, P99	&	$	-12.50	\pm	0.11$	&HAL0968		&	Sakurai's object	\\	
FP J0711-2531 		&	$	-10.69	\pm	0.04	$	&	FBP12	&	$	-10.51	\pm	0.11$	&	HAL0756	&	$\ldots$	\\	
FP J1824-0319	&	$	-10.40	\pm	0.10	$	&	M06, FBP13	&	$	-10.44	\pm	0.10 $	&	HAL1240	&	$\ldots$	\\	
Fr 2-7	&	$	-10.82	\pm	0.10	$	&	FBP13	&	$	-10.86	\pm	0.12$	&	HAL0097	&	$\ldots$	\\	
G4.4+6.4	&	$	-11.25	\pm	0.06	$	&	FBP13	&	$	-11.31	\pm	0.10$	&	HAL0878	&	$\ldots$	\\	
GL1PN J1530-5557 	&	$	-13.86	\pm	0.10	$	&	PC12$^a$&	$	-14.17	\pm	0.12	$&	HAL0233	&	$\ldots$	\\	
GL1PN J1557-5430 	&	$	-14.79	\pm	0.15	$	&	PC12$^a$&	$	-14.70	\pm	0.15$	&	HAL0234	&	$\ldots$	\\	
GL1PN J1642-4453 	&	$	-14.94	\pm	0.15	$	&	PC12$^a$&	$	-14.88	\pm	0.15$	&	HAL0413	&	$\ldots$	\\		
\noalign{\smallskip}
GL1PN J1823-1133	&	$	-13.80	\pm	0.10	$	&	PC12$^a$&	$	-14.05	\pm	0.15	$&	HAL1060	&	$\ldots$	\\	
Hen 2-11  &	$	-10.97	\pm	0.06	$	&	FBP13	&	$	-10.82	\pm	0.08	$&	HAL0459	&	$\ldots$	\\	
HeFa 1	&	$	-12.49	\pm	0.10	$	&	ASTR91	&	$	-12.26	\pm	0.10	$&	HAL0235	&	$\ldots$	\\	%
IC 1295	&	$	-10.65	\pm	0.05	$	&	FBP13	&	$	-10.66	\pm	0.08	$&	HAL1152	&	$\ldots$	\\	%
JaSt 4	&	$	-13.22	\pm	0.10	$	&	JaSt04	&	$	-13.30	\pm	0.10	$&	HAL0709	&	$\ldots$	\\	
JaSt 16	&	$	-14.01	\pm	0.10	$	&	JaSt04	&	$	-14.05	\pm	0.10	$&	HAL0710	&	$\ldots$	\\	
JaSt 17	&	$	-12.74	\pm	0.10	$	&	JaSt04	&	$	-12.71	\pm	0.10$	&	HAL0710	&	$\ldots$	\\	
JaSt 36	&	$	-13.09	\pm	0.10	$	&	JaSt04	&	$	-13.35	\pm	0.10$	&	HAL0710	&	marginally saturated	\\	
JaSt 44	&	$	-13.23	\pm	0.10	$	&	JaSt04	&	$	-13.23	\pm	0.10	$&	HAL0710	&	$\ldots$	\\	
JaSt 63	&	$	-13.95	\pm	0.10	$	&	JaSt04	&	$	-13.57	\pm	0.12	$&	HAL0710	&	$\ldots$	\\	
\noalign{\smallskip}
JaSt 64	&	$	-13.11	\pm	0.03	$	&	JaSt04, RZ04	&	$	-13.22	\pm	0.10	$&	HAL0631	&	$\ldots$	\\	
JaSt 69	&	$	-13.23	\pm	0.10	$	&	JaSt04	&	$	-12.80	\pm	0.10	$&	HAL0710	&	$\ldots$	\\	
JaSt 88	&	$	-12.98	\pm	0.10	$	&	JaSt04	&	$	-12.78	\pm	0.10$	&	HAL0710	&	$\ldots$	\\	
JaSt 95	&	$	-12.54	\pm	0.10	$	&	JaSt04	&	$	-12.54	\pm	0.10	$&	HAL0710	&	$\ldots$	\\	
K 1-4	&	$	-11.65	\pm	0.06	$	&	FBP13	&	$	-11.80	\pm	0.12	$&	HAL0710	&	$\ldots$	\\	
MeWe 1-1	&	$	-11.42	\pm	0.05	$	&	FBP13	&	$	-11.55	\pm	0.09	$&	HAL0177	&	$\ldots$	\\	%
MeWe 1-2	&	$	-11.35	\pm	0.12	$	&	FBP13	&	$	-11.06	\pm	0.08$	&	HAL0222	&	$\ldots$	\\	
MPA J1827-1328 &	$	-12.58	\pm	0.06	$	&	KH93	&	$	-12.71	\pm	0.10	$&	HAL1060	&	$\ldots$	\\
NeVe 3-3	&	$	-12.07	\pm	0.06	$	&	FBP13	&	$	-11.93	\pm	0.10	$&	HAL0600	&	$\ldots$	\\	
NGC 4071 	&	$	-10.95	\pm	0.04	$	&	SK89, FBP13	&	$	-11.11	\pm	0.08$&	HAL0098	&	saturated?	\\	%
\noalign{\smallskip}
PHR J0652-1240 	&	$	-11.64	\pm	0.08	$	&	FBP13	&	$	-11.78	\pm	0.10$	&	HAL1017	&	$\ldots$	\\	
PHR J0719-1222 	&	$	-11.99	\pm	0.10	$	&	FBP13	&	$	-12.03	\pm	0.10$	&	HAL1018	&	$\ldots$	\\	
PHR J0755-3346 	&	$	-12.24	\pm	0.13	$	&	FBP13	&	$	-12.03	\pm	0.10	$&	HAL0600	&	$\ldots$	\\	%
PHR J0808-3745 	&	$	-11.87	\pm	0.12	$	&	FBP13	&	$	-11.81	\pm	0.09$	&	HAL0527	&	$\ldots$	\\	
PHR J0834-2819 	&	$	-11.99	\pm	0.08	$	&	FBP13	&	$	-12.07	\pm	0.08$	&	HAL0679	&	$\ldots$	\\	
PHR J0907-4532 	&	$	-11.58	\pm	0.09	$	&	FBP13	&	$	-11.68	\pm	0.10$	&	HAL0393	&	$\ldots$	\\	
PHR J0941-5356 	&	$	-10.95	\pm	0.07	$	&	FBP13	&	$	-10.85	\pm	0.11$	&	HAL0274	&	$\ldots$	\\	
PHR J0942-5220 	&	$	-11.58	\pm	0.09	$	&	FBP13	&	$	-11.64	\pm	0.10$	&	HAL0274	&	$\ldots$	\\	
PHR J1032-6310 	&	$	-11.40	\pm	0.07	$	&	FBP13	&	$	-11.54	\pm	0.10$	&	HAL0133	&	$\ldots$	\\	
PHR J1052-5042 	&	$	-11.68	\pm	0.06	$	&	FBP13	&	$	-11.51	\pm	0.09$	&	HAL0277	&	$\ldots$	\\	
\noalign{\smallskip}
PHR J1137-6548 	&	$	-10.81	\pm	0.05	$	&	FBP13	&	$	-10.98	\pm	0.11$	&	HAL0134	&	$\ldots$	\\	
PHR J1202-7000 	&	$	-11.80	\pm	0.08	$	&	FBP13	&	$	-11.65	\pm	0.14$	&	HAL0098	&	$\ldots$	\\	
PHR J1246-6324 	&	$	-12.17	\pm	0.09	$	&	FBP13	&	$	-12.37	\pm	0.10$	&	HAL0136	&	$\ldots$	\\	
PHR J1250-6346 	&	$	-11.98	\pm	0.09	$	&	FBP13	&	$	-12.29	\pm	0.10$	&	HAL0136	&	$\ldots$	\\	
PHR J1255-6251 	&	$	-13.00	\pm	0.30	$	&	FBP13	&	$	-12.70	\pm	0.15$	&	HAL0136	&	$\ldots$	\\	
PHR J1315-6555 	&	$	-12.45	\pm	0.02	$	&	PFM11	&	$	-12.82	\pm	0.12$	&	HAL0137	&	$\ldots$	\\	
PHR J1327-6032 	&	$	-11.49	\pm	0.07	$	&	FBP13	&	$	-11.52	\pm	0.12$	&	HAL0180	&	$\ldots$	\\	
PHR J1337-6535 	&	$	-11.56	\pm	0.07	$	&	FBP13	&	$	-11.68	\pm	0.12$	&	HAL0138	&	$\ldots$	\\	
PHR J1408-6106 	&	$	-11.34	\pm	0.07	$	&	FBP13	&	$	-11.46	\pm	0.12$	&	HAL0181	&	$\ldots$	\\	
PHR J1432-6138 	&	$	-11.20	\pm	0.06	$	&	FBP13	&	$	-11.12	\pm	0.15$	&	HAL0182	&	$\ldots$	\\	
\noalign{\smallskip}
PHR J1529-5458 	&	$	-12.47	\pm	0.13	$	&	FBP13	&	$	-12.55	\pm	0.10$	&	HAL0233	&	$\ldots$	\\	
PHR J1537-6159 	&	$	-11.68	\pm	0.13	$	&	FBP13	&	$	-11.59	\pm	0.10$	&	HAL0184	&	$\ldots$	\\	
PHR J1651-3148 	&	$	-11.88	\pm	0.09	$	&	FBP13	&	$	-11.66	\pm	0.10	$&	HAL0628	&	$\ldots$	\\	
PHR J1709-3629	&	$	-12.30	\pm	0.20	$	&	FBP13	&	$	-12.50	\pm	0.09$	&	HAL0554	&	$\ldots$	\\	
PHR J1753-2234	&	$	-12.37	\pm	0.13	$	&	FBP13	&	$	-12.30	\pm	0.11$	&	HAL0794	&	$\ldots$	\\	
PHR J1806-1956 	&	$	-12.40	\pm	0.20	$	&	FBP13	&	$	-12.41	\pm	0.12$	&	HAL0880	&	$\ldots$	\\	
PHR J1818-1526 	&	$	-12.80	\pm	0.20	$	&	FBP13	&	$	-12.77	\pm	0.12$	&	HAL0970	&	$\ldots$	\\	
PM 1-104	&	$	-13.50	\pm	0.10	$	&	PC12	&	$	-13.37	\pm	0.12	$&	HAL0234	&	$\ldots$	\\	
PTB 15	&	$	-12.44	\pm	0.10	$	&	B03	&	$	-12.34	\pm	0.10$	&	HAL0968&	$\ldots$	\\	%
PTB 17	&	$	-12.15	\pm	0.03	$	&	B03, FBP13	&	$	-12.35	\pm	0.10	$& HAL0968		&	$\ldots$	\\	%
\noalign{\smallskip}
PTB 23	&	$	-11.80	\pm	0.20	$	&	FBP13	&	$	-11.59	\pm	0.08$	&	HAL0971	&	$\ldots$	\\	
PTB 24	&	$	-13.06	\pm	0.20	$	&	B03	&	$	-12.99	\pm	0.09	$&	HAL1059	&	$\ldots$	\\	
PTB 30	&	$	-12.95	\pm	0.20	$	&	B06	&	$	-12.85	\pm	0.10$	&	HAL0879	&	$\ldots$	\\	
RCW 24	&	$	-10.90	\pm	0.04	$	&	FBP13	&	$	-10.80	\pm	0.12$	&	HAL0458	&	$\ldots$	\\	
Sab 41	&	$	-11.41	\pm	0.05	$	&	FBP13	&	$	-11.34	\pm	0.11	$&	HAL0556	&	$\ldots$	\\	%
SaWe 3	&	$	-11.40	\pm	0.09	$	&	HDM98, FBP13	&	$	-11.42	\pm	0.12	$&	HAL0133	&	$\ldots$	\\	%
SB 3	&	$	-11.63	\pm	0.07	$	&	BDF99	&	$	-11.63	\pm	0.09	$&	HAL0632	&	$\ldots$	\\	
SB 8	&	$	-12.25	\pm	0.12	$	&	BDF99, FBP13	&	$	-12.28	\pm	0.10	$&	HAL0712	&	$\ldots$	\\	
SB 12	&	$	-13.65	\pm	0.10	$	&	BDF99	&	$	-13.77	\pm	0.10	$&	HAL0712	&	$\ldots$	\\	
SB 33	&	$	-13.05	\pm	0.10	$	&	BDF99	&	$	-12.75	\pm	0.10	$&	HAL0485	&	$\ldots$	\\	%
\noalign{\smallskip}
SB 34	&	$	-12.57	\pm	0.10	$	&	BDF99	&	$	-12.60	\pm	0.09	$&	HAL0485	&	$\ldots$	\\	%
Sh 2-42	&	$	-11.04	\pm	0.05	$	&	FBP13	&	$	-10.92	\pm	0.10$	&	HAL0969	&	$\ldots$	\\	
Sh 2-68 &	$	-10.46	\pm	0.08	$	&	XPPT, FBP13 &   	$	-10.49	\pm	0.10	$&	HAL1330	&	uncertain PN	\\	
SuWt 2	&	$	-11.50	\pm	0.20	$	&	FBP13	&	$	-11.70	\pm	0.15$	&	HAL0181	&	$\ldots$	\\	%
vBe 2	&	$	-12.27	\pm	0.13	$	&	FBP13	&	$	-12.24	\pm	0.12	$&	HAL0234	&	$\ldots$	\\	
WeSb 4	&	$	-12.31	\pm	0.15	$	&	FBP13	&	$	-12.37	\pm	0.13$	&	HAL1332	&	$\ldots$ \\	
WKG 1	&	$	-12.13	\pm	0.10	$	&	WKG97	&	$	-12.37	\pm	0.10$	&	HAL0138	&	$\ldots$   \\	
WKG 3	&	$	-12.26	\pm	0.12	$	&	FBP13	&	$	-12.30	\pm	0.10	$     &	HAL0181	&	$\ldots$	\\	
\hline																			
Hf 39		&	$	-11.74	\pm	0.10	$	&	S87, H94	&	$	-11.52	\pm	0.10	$&	HAL0175	&	LBV ejecta	\\	
K 2-15			&	$	-10.41	\pm	0.04	$	&	FBP13	&	$	-10.45	\pm	0.08	$&	HAL0392	&	\HII\ region	\\	
NGC 2736		&	$	-10.27	\pm	0.07	$	&	this work	&	$	-10.39	\pm	0.10$	&	HAL0392	&	part of Vela SNR	\\	
RCW 27			&	$	-7.64	\pm	0.06	$	&	this work	&	$	-7.67	\pm	0.08	$&		HAL0459				&	\HII\  region	\\	
RCW 32			&	$	-8.60	\pm	0.07	$	&	this work	&	$	-8.55	\pm	0.07	$&		HAL0459				&	\HII\ region 	\\	
RCW 33			&	$	-7.71\pm	0.08	$	&	this work&	$	-7.66	\pm	0.08$	&	HAL0459					&	\HII\  region	\\	
RCW 36			&	$	-9.34	\pm	0.05	$	&	this work	&	$	-9.47	\pm	0.10$	&	HAL0392	&	\HII\  region	\\	
RCW 40			&	$	-9.02	\pm	0.05	$	&	this work	&	$	-9.11	\pm	0.10$	&	HAL0331	&	\HII\ region 	\\	
RCW 58			&	$	-9.70	\pm	0.06	$	&     this work&	$	-10.11	\pm	0.12	$&	HAL0133	&	WR ejecta	\\	
RCW 64			&	$	-10.24	\pm	0.05	$	&	this work&	$	-10.29	\pm	0.08	$&	HAL0135	&	\HII\    region	\\	
\noalign{\smallskip}
Sh 2-61	&	$	-10.02	\pm	0.06	$	&	FBP13	&	$	-10.20	\pm	0.10$	&	HAL0350	&	HII	region\\	
vBe 1	&	$	-10.02	\pm	0.06	$	&	FBP13	&	$	-10.20	\pm	0.10$	&	HAL0350	&	HII	region\\	
WR 16	&	$	-10.64	\pm	0.10	$	&	this work&	$		-10.57\pm	 0.10$	&	HAL0221					&	WR shell	\\	
Wray 15-751	&	$	-12.43	\pm	0.08	$	&	H94	&	$	-12.25	\pm 0.09	$	&	HAL0176					&	LBV ejecta	\\	
\end{longtable}
\flushleft
\noindent{Reference codes:  ASTR91 -- Acker et al. (1991);   B03, B06 -- Boumis et al. (2003, 2006); BDF99 -- Beaulieu et al. (1999); DB96 -- Duerbeck \& Benetti (1996);  FBP13 --  Frew et al. (2013); FBP14 --  Frew et al. (2014);  H94 -- Hutsemekers (1994); HDM98 -- Hua, Dopita \& Martinis (1998); JaSt04 -- Jacoby \& Van de Steene (2004); KH93 -- Kistiakowsky \& Helfand (1993);  M06 -- Madsen et al. (2006b); P99 -- Pollacco (1999);  PC12 -- Parker et al. (2012b);  PFM11 -- Parker et al. (2011); RZ04 -- Ruffle et al. (2004);  S87 -- Stahl (1987);  SK89 -- Shaw \& Kaler (1989); WKG97 -- Weinberger et al. (1997); XP94 -- Xilouris et al. (1994);  XPPT -- Xilouris et al. (1996). ~~Note:~$^a$ fluxes revised using new calibration factors from this work.  }
\end{center}
}

\end{document}